\documentclass[a4paper, 12pt]{article}

\pdfoutput=1
\usepackage{cite}
\usepackage[affil-it]{authblk}
\usepackage[para]{footmisc}
\usepackage{sectsty}
\sectionfont{\large}
\subsectionfont{\normalsize}
\subsubsectionfont{\normalsize}
\paragraphfont{\normalsize}

\usepackage{cmap}
\usepackage{amsmath}
\usepackage{amssymb}
\usepackage{mathtools}
\usepackage{graphicx}
\usepackage{fancyhdr}
\usepackage{indentfirst}
\usepackage{array}
\usepackage{subcaption}
\usepackage{booktabs}
\usepackage[dvipsnames]{xcolor}
\usepackage{hyperref}
\hypersetup{
    colorlinks=true,
    linkcolor={red!50!black},
    citecolor={blue!80!black},
    urlcolor={blue!80!black},
    pdftitle={On the layer crossing problem for a semi-infinite hydraulic fracture},
    pdfauthor={A.V. Valov, E.V. Dontsov}
}

\newcommand{\opd}[2]{ \frac{\mathrm{d} #1}{\mathrm{d} #2} }
\newcommand{\norm}[1]{\left\lVert#1\right\rVert}
\newcommand{\dint}{\,\mathrm{d}}
\DeclareMathOperator*{\argmin}{arg\,min}
\newcommand*\widebar[1]{\overline{#1}}

\newcommand{\tildeW}{\tilde{w}}            % tilde w
\newcommand{\tildeS}{\tilde{s}}            % tilde s

\newcommand{\Eprime}{E'}           % E'
\newcommand{\Kprime}{K'}           % K'
\newcommand{\Cprime}{C'}           % C'
\newcommand{\muprime}{\mu'}        % mu'

\begin{document}

\title{\Large On the layer crossing problem for a semi-infinite hydraulic fracture}

\author[1, 2]{\normalsize A.V. Valov\thanks{a.valov@g.nsu.ru}}
\author[3]{\normalsize E.V. Dontsov\thanks{egor@resfrac.com}}
\affil[1]{\footnotesize Lavrentyev Institute of Hydrodynamics SB RAS, Novosibirsk, 630090, Russia}
\affil[2]{\footnotesize Novosibirsk State University, Novosibirsk, 630090, Russia}
\affil[3]{\footnotesize ResFrac Corporation, Palo Alto, CA 94301, USA}

\date{}
\maketitle

\begin{abstract}
    This paper analyses the problem of a semi-infinite fluid-driven fracture propagating through multiple stress layers in a permeable elastic medium. Such a problem represents the tip region of a planar hydraulic fracture. When the hydraulic fracture crosses a stress layer, the use of a standard tip asymptotic solution may lead to a considerable reduction of accuracy, even for the simplest case of a height-contained fracture. In this study, we propose three approaches to incorporate the effect of stress layers into the tip asymptote: non-singular integral formulation, toughness-corrected asymptote, and an ordinary differential equation approximation of the non-singular integral formulation mentioned above. As illustrated in the paper, these approaches for stress-corrected asymptotes differ in computational complexity, the complexity of implementation, and the accuracy of the approximation. In addition, the size of the validity region of the stress-corrected asymptote is evaluated, and it is shown to be greatly reduced relative to the case without layers. In order to address the issue, the stress relaxation factor is introduced. This, in turn, allows for enhancing the accuracy of the layer-crossing computation on a relatively coarse mesh to utilize the stress-corrected asymptote in hydraulic fracturing simulators for the purpose of front tracking.
\end{abstract}

\section{Introduction}
    Hydraulic fractures are a specific class of tensile fractures propagating underground in pre-stressed solid media due to high-pressure injection of viscous fluid. These fractures are commonly used in the oil and gas industry to increase hydrocarbon recovery, primarily for unconventional reservoirs~\cite{Economides_Reservoir_Stimulation_1989}. Hydraulic fractures also occur naturally as kilometer-long dikes driven by magma from deep underground chambers to the earth’s surface~\cite{Rubin_magma_filled_cracks_1995, Roper_buoyancy_driven_cracks_2007, Rivalta_dikes_models_review_2015}. In the mining industry, hydraulic fractures are used to precondition the rock mass with the aim of improving caveability and fragmentation for block-caving mining operations~\cite{Jeffrey_hydraulic_fracturing_mining_2000, Katsaga_hydraulic_fracturing_mining_2015}. In addition, hydraulic fractures are used for accelerating the waste remediation process~\cite{Frank_remediation_by_fracturing_1995}, waste disposal~\cite{Abou_oily_waste_injection_1989}, and enhancing the injectivity and capacity of geologic carbon storage reservoirs~\cite{Huerta_hydraulic_fracturing_CO2_2020}.

    Mathematical models for hydraulic fracturing involve the non-local hypersingular integral equation governing the elastic response, the non-linear and history-dependent equation of the flow of a viscous fluid inside the fracture, and a free boundary problem determined by fracture propagation. The asymptotic behavior in the vicinity of the fracture tip has a pivotal influence on fracture propagation~\cite{Spence_Sharp_self_similar_1985, Desroches_crack_tip_region_1994, Detournay_Propagation_regimes_2004, Garagash_Detournay_Tip_Region_of_a_Fluid_Driven_2000}. The hydraulic fracture propagation condition for rocks is typically based on the classical square-root crack tip singularity stemming from the Linear Elastic Fracture Mechanics (LEFM). However, an analysis of the tip region shows that the LEFM square-root asymptote is often limited to a small domain near the fracture tip~\cite{Garagash_Detournay_plane_strain_small_toughness_2005, Garagash_Multiscale_Tip_Asymptotic_2011}. Numerous studies have addressed the problem of asymptotic solutions beyond the square-root solution. The papers~\cite{Garagash_Detournay_Tip_Region_of_a_Fluid_Driven_2000, Garagash_Detournay_plane_strain_small_toughness_2005, Mitchell_Peirce_asymptotic_framework_2007} have identified a multiscale nature of the tip region problem when multiple physical processes compete to control the global response of fluid-driven fractures. Recently, studies~\cite{Garagash_Multiscale_Tip_Asymptotic_2011, Dontsov_Peirce_universal_asymptotic_2015} proposed the three-processes tip asymptote that simultaneously captures fracture toughness, fluid viscosity, and leak-of. The paper~\cite{Dontsov_Peirce_ILSA_2017} provides an approximate implicit closed-form solution for the multiscale tip problem that is suitable for implementation in a hydraulic fracturing simulator. Many other studies consider a variety of effects, such as non-Newtonian fluid rheology~\cite{Dontsov_Kresse_2018, Moukhtari_Lecampion_semi_infinite_non_newtonian_2018}, consideration of the poroelastic stress influence~\cite{Kovalyshen_semi_infinite_poroelastic_2013}, hydraulic fracture with pressure-dependent leak-off~\cite{Kanin_radial_pdl_2020}, and studying deflating hydraulic fracture~\cite{Peirce_Detournay_deflating_asymptote_2022}. 

    In advanced hydraulic fracturing simulators, such as Implicit Level Set Algorithm (ILSA)~\cite{Dontsov_Peirce_ILSA_2017}, utilizing the asymptotic solution without a built-in layer-crossing logic may lead to a significant loss of accuracy when the fracture crosses the boundary between layers with different confining stresses~\cite{Dontsov_Front_Tracking_Comparison_2022}. There are several studies that attempted to address the problem. The paper~\cite{Dontsov_Homogenization_2017} considers the numerical solution to the problem of a hydraulic fracture propagating through multiple stress layers under plane-strain elastic conditions. The solution in the tip region is based on a non-singular integral formulation for a semi-infinite hydraulic fracture that precisely accounts for the variation of the confining stress near the tip. A different study developed an algorithm that uses Multi Layer Tip Elements (MuLTipEl)~\cite{Dontsov_MuLTipEl_2022, Dontsov_MuLTipEl_SPE_2022}. The MuLTipEl approach proposes a different front-tracking algorithm from ILSA, but still utilizes the near-tip asymptotic solution for the purpose of front tracking. In addition, it incorporates variations of the confining stress, leak-off coefficient, and fracture toughness from thin layers that can be smaller than the element size. In view of these results, this study aims to further investigate the problem of a fracture tip propagating through multiple stress layers and to develop an algorithm that is capable of capturing the effects of stress layering and the multiscale nature of the solution in the tip region. In contrast to the aforementioned approaches, the goal is to develop an algorithm that is more computationally efficient than rigorously solving for the tip asymptote numerically, as was done in~\cite{Dontsov_Homogenization_2017}. At the same time, the approach should be more accurate than the approximation used in MuLTipEl~\cite{Dontsov_MuLTipEl_2022, Dontsov_MuLTipEl_SPE_2022}.

    The paper is organized as follows. Section~\ref{sec:semi_infinite_fracture} describes the problem formulation and the governing equations for the semi-infinite hydraulic fracture propagating in a homogeneous medium, i.e. without stress layers. Section~\ref{sec:layer_crossing_problem} provides motivation for using a tip asymptotic solution accounting for the effect of stress layers. After that, Section~\ref{sec:tip_asymptotic_solutions} introduces three various approaches to include the effect of stress layers. Section~\ref{sec:comparison_of_approaches} presents a comparative analysis of the accuracy and computational complexity of the proposed approaches. Finally, Section~\ref{sec:validity_region} quantifies the size of the tip region, in which the asymptotic solution with stress layers remains accurate, and proposes a method to extend the validity region.

\section{A semi-infinite hydraulic fracture in a homogeneous medium}\label{sec:semi_infinite_fracture}
\begin{figure}
        \centering
        \includegraphics[width=0.5\linewidth]{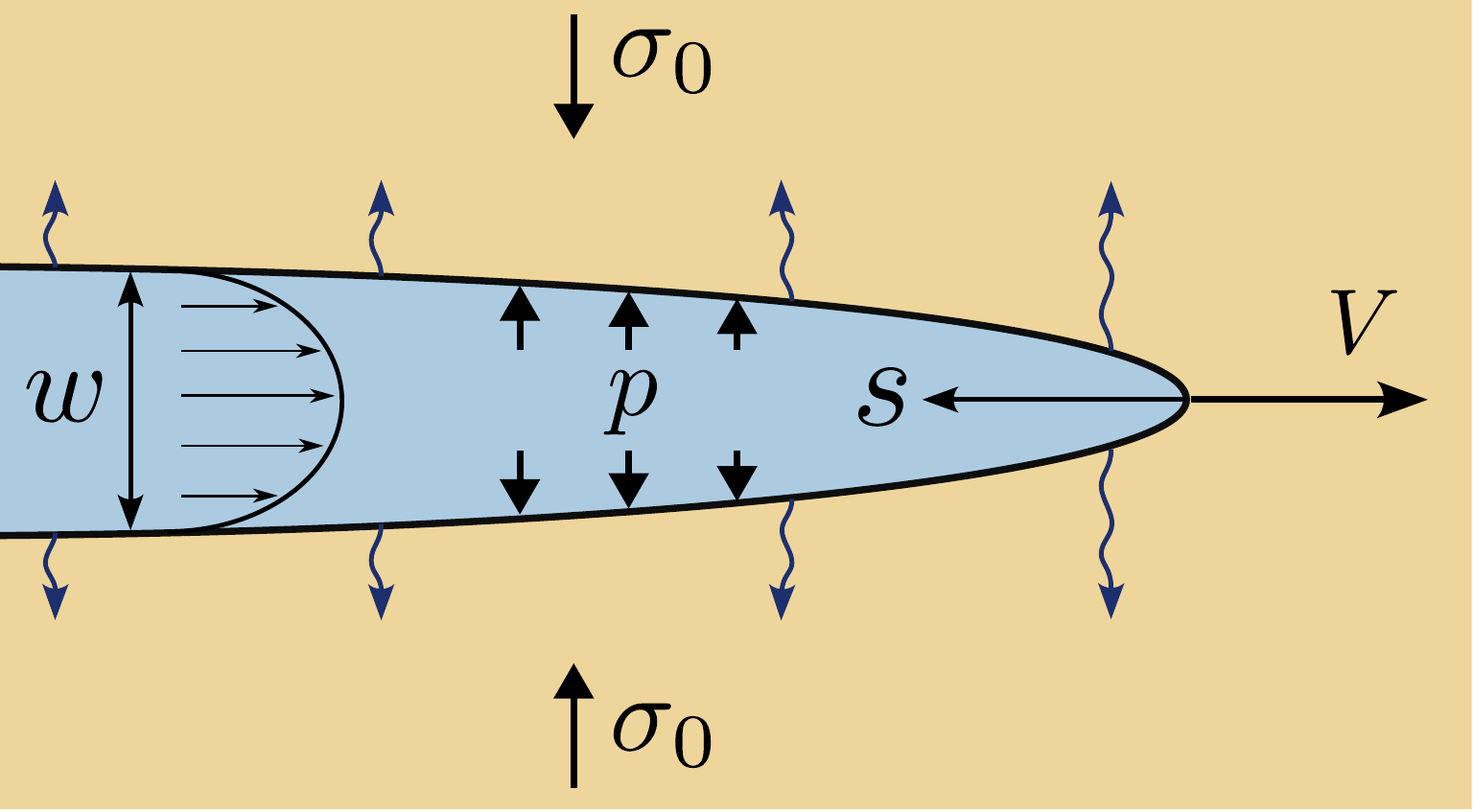}
        \caption{Schematics for a semi-infinite hydraulic fracture propagating in a homogeneous formation at a constant velocity $V$.}
        \label{img:semi_infinite_fracture}
    \end{figure}

    As shown in~\cite{Peirce_Detournay_ILSA_2008, Garagash_Multiscale_Tip_Asymptotic_2011}, the solution near the tip of a planar hydraulic fracture can be approximated by the problem of a semi-infinite hydraulic fracture propagating at a constant velocity $V$ under plain strain elastic conditions. Note that the propagation velocity $V$ is the instantaneous tip velocity that is provided for the semi-infinite fracture problem. It is convenient to introduce a moving coordinate $s$ with the origin at the fracture tip so that $s$ quantifies the distance from a point inside the fracture to the fracture tip, see Figure~\ref{img:semi_infinite_fracture}.

    To simplify further mathematical expressions, let us introduce the following scaled parameters
    \begin{equation}\label{scaled_parameters}
        \Eprime = \frac{E}{1 - \nu^2}, \qquad \Kprime = \sqrt{\frac{32}{\pi}} K_\mathrm{Ic}, \qquad \Cprime = 2C_L, \qquad \muprime = 12\mu,
    \end{equation}
    where $E$ is the Young's modulus, $\nu$ is the Poisson's ratio, $K_\mathrm{Ic}$ is the fracture toughness, $C_L$ is the Carter's leak-off coefficient, and $\mu$ is the fluid viscosity. These parameters are assumed to be constant, i.e. the rock formation is assumed to have homogeneous elastic properties, toughness, and leak-off, while a single fluid type is injected to drive the fracture propagation.

    The LEFM propagation condition~\cite{Rice_Math_Fracture_1968} for a pure mode I fracture $K_\mathrm{I} = K_\mathrm{Ic}$ can be expressed as an asymptotic behavior of the fracture width $w$ in the vicinity of the fracture tip
    \begin{equation}\label{semi_infinite_propagation_cond}
        w = \frac{\Kprime}{\Eprime} s^{1/2}, \quad s \rightarrow 0.
    \end{equation}
    The lubrication equation for an incompressible fluid in the moving coordinate system $s$ reduces to~\cite{Peirce_Detournay_ILSA_2008, Garagash_Multiscale_Tip_Asymptotic_2011, Dontsov_Peirce_universal_asymptotic_2015}
    \begin{equation}\label{semi_infinite_lubrication}
        \frac{w^2}{\muprime} \opd{p}{s} = V + 2\Cprime V^{1/2} \frac{s^{1/2}}{w},
    \end{equation}
    where $w$ is the fracture width and $p$ is the fluid pressure. According to~\cite{Rice_Math_Fracture_1968}, the elastic relation between the net pressure $p(s) - \sigma_0$ and the fracture width $w(s)$ can be expressed as the following integral equation
    \begin{equation}\label{semi_infinite_elasticity}
        p(s) - \sigma_0 = \frac{\Eprime}{4\pi}\int_{0}^{\infty} \opd{w(x')}{x'}\frac{\dint x'}{s - x'}.
    \end{equation}
    An inversion of the elasticity equation~\eqref{semi_infinite_elasticity} that also includes the propagation condition~\eqref{semi_infinite_propagation_cond} takes the following form~\cite{Garagash_Detournay_Tip_Region_of_a_Fluid_Driven_2000, Dontsov_Peirce_universal_asymptotic_2015}
    \begin{equation}\label{semi_infinite_elasticity_inversed}
        w(s) = \frac{\Kprime}{\Eprime} s^{1/2} - \frac{4}{\pi\Eprime}\int_{0}^{\infty} F(s, x') \opd{(p - \sigma_0)}{x'} \dint x',
    \end{equation}
    where the kernel $F(s, x')$ is given by
    \begin{equation}\label{semi_infinite_kernel_F}
        F(s, x') = (x' - s) \ln \left|\frac{s^{1/2} + {x'}^{1/2}}{s^{1/2} - {x'}^{1/2}}\right| - 2 s^{1/2} {x'}^{1/2}.
    \end{equation}
    Consistent with~\cite{Peirce_Detournay_ILSA_2008, Dontsov_Peirce_universal_asymptotic_2015}, in this section we assume that the far-field confining stress $\sigma_0$ is spatially constant, which implies that it does not affect the solution.

    A combination of the lubrication equation~\eqref{semi_infinite_lubrication} and LEFM asymptotic fracture width~\eqref{semi_infinite_propagation_cond} implies that the fluid pressure has a logarithmic singularity at the fracture tip. To mitigate the problem, a so-called non-singular integral formulation is employed~\cite{Dontsov_Peirce_universal_asymptotic_2015}. The pressure gradient from the lubrication equation~\eqref{semi_infinite_lubrication} is substituted into the inverted elasticity equation~\eqref{semi_infinite_elasticity_inversed} to obtain
    \begin{equation}\label{semi_infinite_integral_equation}
        w(s) = \frac{\Kprime}{\Eprime} s^{1/2} - \frac{4}{\pi\Eprime} \int_{0}^{\infty} F(s, x')\frac{\muprime}{w(x')^2} \left[V + 2\Cprime V^{1/2}\frac{{x'}^{1/2}}{w(x')}\right] \dint x'.
    \end{equation}
    The above equation simplifies the original problem since the singular pressure is eliminated from the solution, while the equation~\eqref{semi_infinite_integral_equation} is an integral formulation written only for the fracture width $w(s)$.

    By introducing the following scaled quantities~\cite{Dontsov_Homogenization_2017}
    \begin{equation}\label{semi_infinite_scaling}
        \begin{gathered}
            \tilde{w} = \frac{\Eprime w}{\Kprime s^{1/2}}, \quad 
            \chi = \frac{2 \Cprime \Eprime}{V^{1/2} \Kprime}, \quad
            l = \left(\frac{\Kprime^3}{\muprime \Eprime^2 V}\right)^2, \\
            \tilde{s} = \left(\frac{s}{l}\right)^{1/2}, \quad 
            \tilde{x}' = \left(\frac{x'}{l}\right)^{1/2},
        \end{gathered}
    \end{equation}
    equation~\eqref{semi_infinite_integral_equation} can be further reduced to
    \begin{equation}\label{semi_infinite_integral_equation_dimless}
        \tildeW(\tildeS) = 1 + \frac{8}{\pi} \int_{0}^{\infty} G\left(\frac{\tilde{x}'}{\tildeS}\right) \left[\frac{1}{\tildeW(\tilde{x}')^2} + \frac{\chi}{\tildeW(\tilde{x}')^3}\right] \dint\tilde{x}',
    \end{equation}
    where the new non-singular kernel is
    \begin{equation}
        G(t) = \frac{1 - t^2}{t}\ln\left|\frac{1 + t}{1 - t}\right| + 2.
    \end{equation}
    The integral equation~\eqref{semi_infinite_integral_equation_dimless} is the non-singular formulation for the problem.

    The problem of a semi-infinite fracture propagating in a homogeneous formation has three limiting regimes of propagation: toughness dominated ($k$), leak-off dominated ($\tilde{m}$), and viscosity dominated ($m$). In particular, the $k$-regime corresponds to the situation in which the energy dissipation for a new fracture surface creation dominates while the effects of fluid viscosity and leak-off are negligible. The $m$-regime is characterized by the dominant role of viscous energy dissipation, while toughness and leak-off can be neglected. At the same time, the leak-off dominated regime $\tilde m$ corresponds to the high leak-off case. For a detailed asymptotic analysis of a semi-infinite fracture in a homogeneous formation, we refer to~\cite{Garagash_Detournay_Tip_Region_of_a_Fluid_Driven_2000, Detournay_Propagation_regimes_2004, Garagash_Multiscale_Tip_Asymptotic_2011}. The three terms on the right-hand side of the integral equation~\eqref{semi_infinite_integral_equation_dimless} correspond to these limiting regimes. By substituting solution in the form of $\tildeW = \beta \tildeS^{\delta}$ into~\eqref{semi_infinite_integral_equation_dimless} and keeping only one term that corresponds to either toughness, viscosity, or leak-off, one finds that
    \begin{equation}\label{vertex_solutions_homogeneous}
        \tildeW_{k} = 1, \quad \tildeW_{\tilde{m}} = \beta_{\tilde{m}} \chi^{1/4} \tildeS^{1/4}, \quad \tildeW_{m} = \beta_{m} \tildeS^{1/3},
    \end{equation}
    where $\beta_m = 2^{1/3} 3^{5/6}$, $\beta_{\tilde{m}} = 4 / (15 (\sqrt{2} - 1))^{1/4}$, and $\tildeW_{k}$ corresponds to the toughness dominated solution, $\tildeW_{\tilde{m}}$ is the leak-off dominated solution, while $\tildeW_{m}$ is the viscosity dominated solution. The complete solution transitions from one limiting solution to another depending on problem parameters.
    
\begin{figure}
        \centering
        \includegraphics[width=0.7\linewidth]{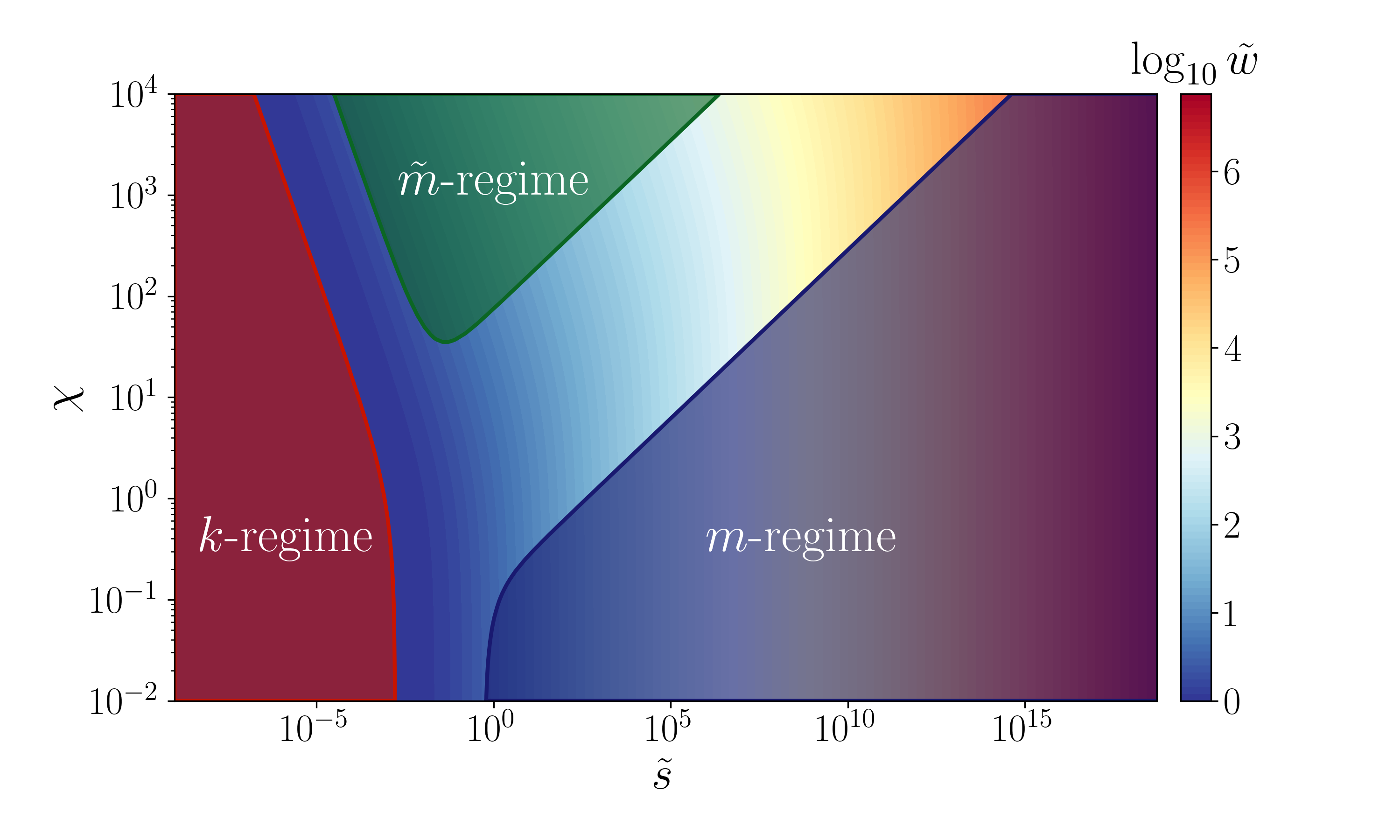}
        \caption{Parametric space for a semi-infinite hydraulic fracture. Regions of applicability of the limiting regimes are highlighted by the red ($k$-regime or toughness dominated), the green ($\tilde{m}$-regime or leak-off dominated), and the blue ($m$-regime or viscosity dominated) colors.}
        \label{img:regimes_map}
    \end{figure}

    According to~\cite{Garagash_Multiscale_Tip_Asymptotic_2011, Dontsov_Peirce_universal_asymptotic_2015}, the propagation regime for a semi-infinite fracture is governed by the leak-off parameter $\chi$ and the scaled distance $\tildeS$. Figure~\ref{img:regimes_map} shows the parametric space for a semi-infinite fracture and the scaled fracture width $\tilde{w}$ that is the solution of the integral equation~\eqref{semi_infinite_integral_equation_dimless}. The regions of applicability of the limiting regimes~\eqref{vertex_solutions_homogeneous} and their dependence on the leak-off parameter $\chi$ and the scaled distance $\tilde{s}$ are highlighted by the red ($k$-regime or toughness dominated), the green ($\tilde{m}$-regime or leak-off dominated), and the blue ($m$-regime or viscosity dominated) colors. The regions of applicability are defined as the regions where the difference of the solution of the integral equation~\eqref{semi_infinite_integral_equation_dimless} with corresponding limiting regimes~\eqref{vertex_solutions_homogeneous} is below 2 percent. 

    The summary of the universal asymptotic solution presented in this section covers a baseline solution without layers. As will be shown in next section, such a solution leads to numerical artifacts when fracture crosses a stress layer and therefore the effect of stress layers needs to be included. At the same time, all the updated solutions that account for the effect of stress layers need to accurately recover this layer-free solution in the case of zero stress change between the layers.

\section{Motivation for the layer crossing problem}\label{sec:layer_crossing_problem}
    Advanced hydraulic fracturing simulators, such as Implicit Level Set Algorithm (ILSA)~\cite{Peirce_Detournay_ILSA_2008, Dontsov_Peirce_ILSA_2017}, employ the so-called tip asymptotic solutions to track the location of the fracture front. This leads to a more accurate representation of fracture geometry and increases the overall accuracy of the solution. However, as will be shown in this section, the approach loses some of its advantages when the fracture crosses a stress layer.

\begin{figure}
        \centering
        \includegraphics[width=0.9\linewidth]{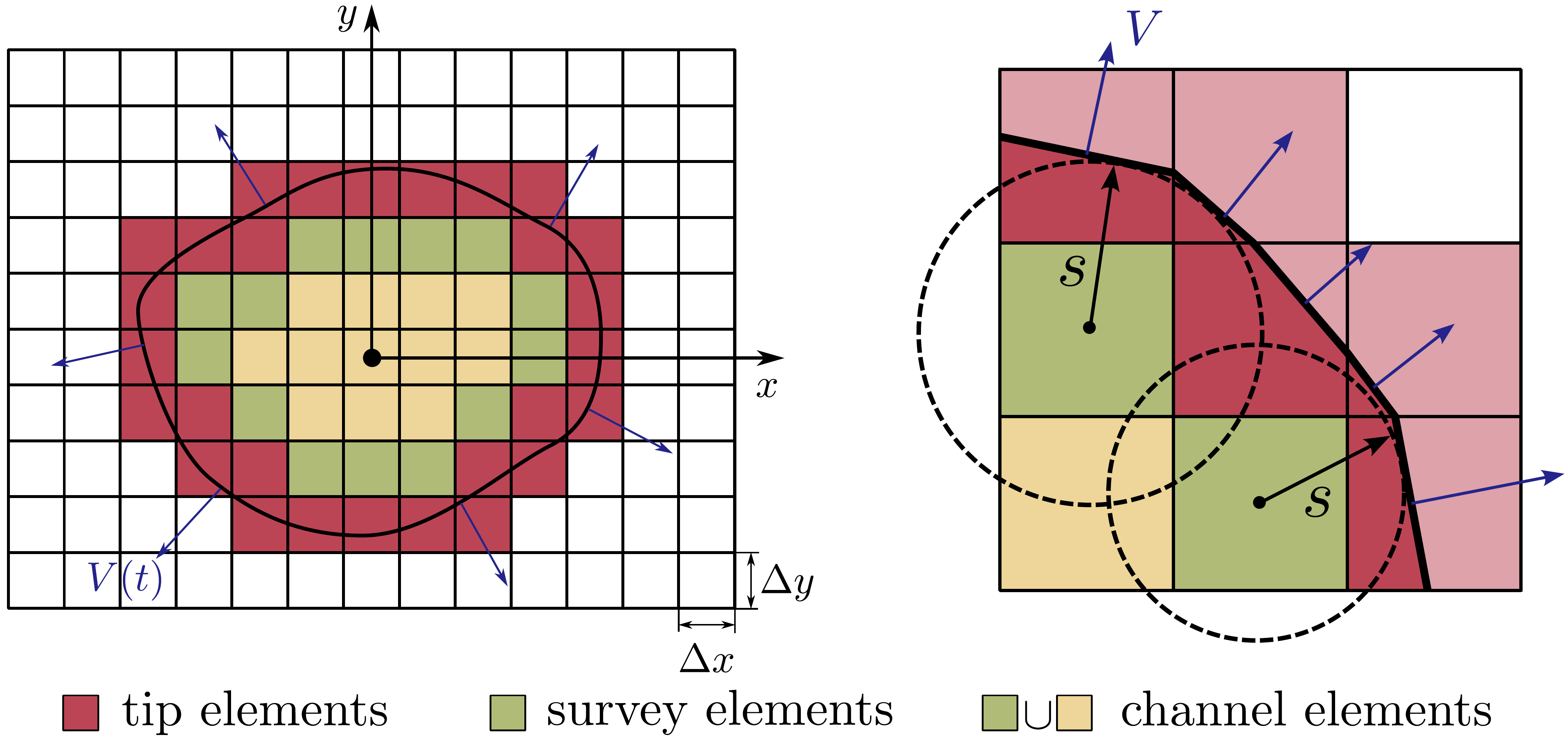}
        \caption{Left panel: Discretization for a planar fracture using a fixed rectangular mesh. Classification of the fracture elements into channel, tip, and survey. The black line shows the fracture front and the arrows schematically indicate the direction of propagation. Right panel: schematics of the fracture front tracking and tip volume calculation, where $s$ is the distance from the center of the survey element to the fracture front. The fluid-filled part of the tip element is highlighted by the dark red color, while the remaining unfilled regions of the tip element are shown by the light red color.}
        \label{img:descrete_mesh}
    \end{figure}

    To this end, a brief overview of the ILSA front tracking methodology is given for the case of a single planar hydraulic fracture. The fracture area is covered with rectangular elements, which are divided into three types: channel, tip, and survey, see Figure~\ref{img:descrete_mesh}. The channel elements are fully contained within the fracture footprint. The tip elements are partially filled with fluid and intersected by the fracture front. The survey elements are a subset of the channel elements and have at least one adjacent tip element. Survey elements are utilized to locate the fracture front using asymptotic solutions.
    
    The ILSA approach invokes the use of the tip asymptotic solution to determine the distance to the fracture front. One possibility is to use the so-called universal asymptote, which is the solution of the integral equation~\eqref{semi_infinite_integral_equation}. The approximate solution is presented in~\cite{Dontsov_Peirce_universal_asymptotic_2015} and was applied to ILSA in~\cite{Dontsov_Peirce_ILSA_2017}. As was mentioned in the previous section, this solution simultaneously captures the effects of rock toughness, fluid viscosity, and leak-off in the tip region and can be expressed as
    \begin{equation}\label{universal_asymptotic}
        \frac{s^{2} V \mu'}{E'w^{3}} = g_\delta \left(\frac{K' s^{1/2}}{E' w}, \frac{2s^{1/2} C'}{w V^{1/2}} \right),
    \end{equation}
    where $V$ is the fracture front velocity, $w(s)$ is the tip asymptotic solution, $s$ is the distance to the fracture front, while the definition of the function $g_\delta$ is given in~\cite{Dontsov_Peirce_ILSA_2017}.

    For the given fracture width $w^s$ at the center of a survey element, the distance to the fracture front $s$ can be calculated by solving the nonlinear equation~\eqref{universal_asymptotic} with $w = w^s$. The fracture front velocity is approximated as $V = (s - s_0) / \Delta t$, where $s_0$ is the distance from the center of the survey element to the fracture front at the previous time step. The distance $s$ is calculated for each survey element, and it is used to locate the fracture front as shown in the right panel of Figure~\ref{img:descrete_mesh}. In addition, the fluid in the tip element occupies a polygonal region highlighted by the dark red color. In contrast, the unfilled part of the tip element is highlighted by the light red color. For the purpose of the numerical scheme, the width of the partially filled tip element $w^t$ is calculated from the volume of fluid contained in this element as
    \begin{equation}
        w^t = \frac{1}{\Delta x \Delta y} \int_{PR} w(x, y) \dint x \dint y,
    \end{equation}
    where $PR$ is the polygonal region of the tip element occupied by the fluid. An effective procedure for integrating the asymptotic solution $w$ over the tip element is described in detail in~\cite{Dontsov_Peirce_ILSA_2017}. One important feature of the algorithm is that the widths of the tip elements are not independent and are uniquely expressed in terms of the fracture openings of the survey elements.
    
\begin{figure}
        \centering
        \begin{subfigure}{0.99\textwidth}
            \centering
            \includegraphics[width=0.8\linewidth]{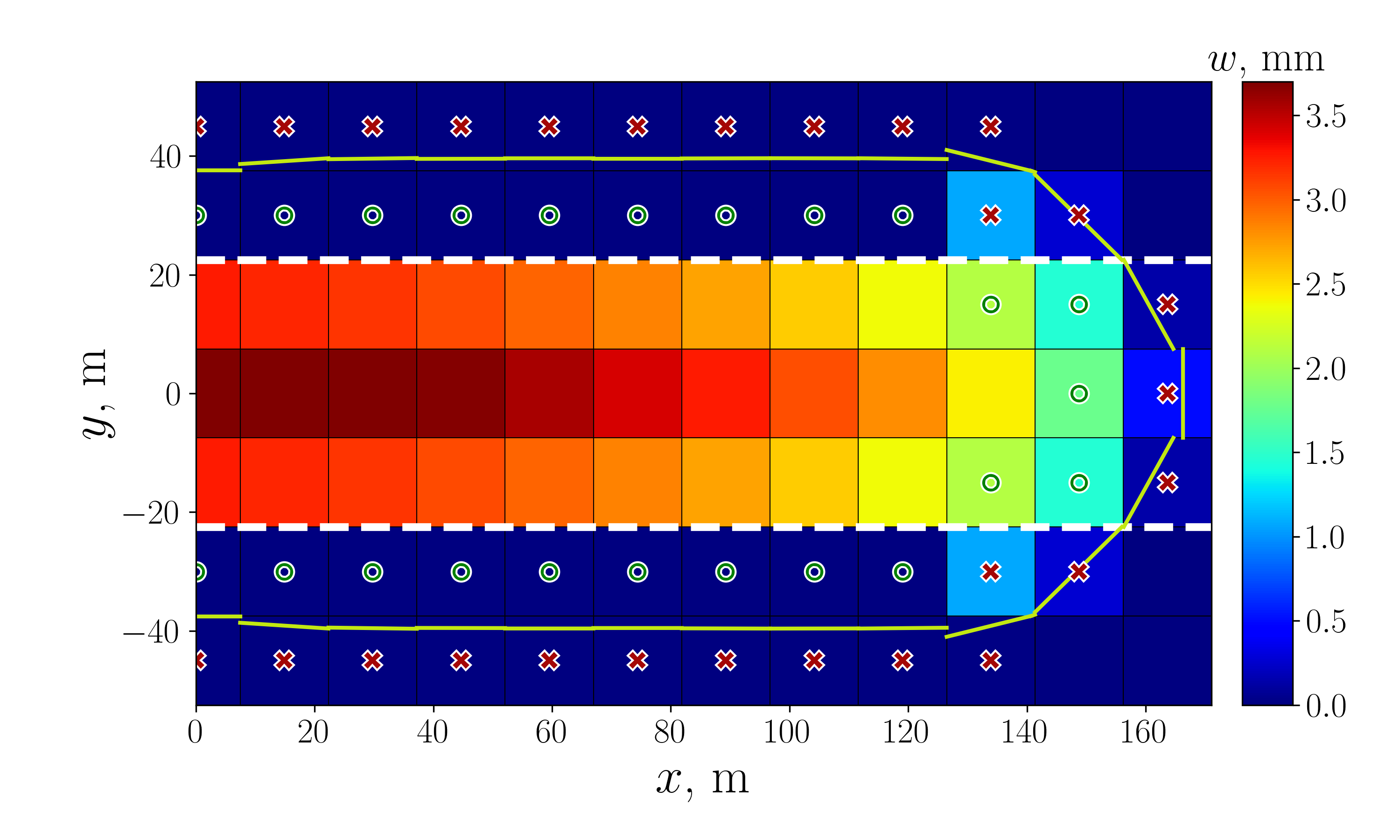}
            \caption{Example numerical calculation for a planar hydraulic fracture for the case of strong symmetric stress barriers. The color filling indicates the fracture opening, while the yellow line shows the fracture front.}
            \label{img:horns_problem:calculation}
        \end{subfigure}
        \\
        \begin{subfigure}{0.99\textwidth}
            \centering
            \includegraphics[width=1.0\linewidth]{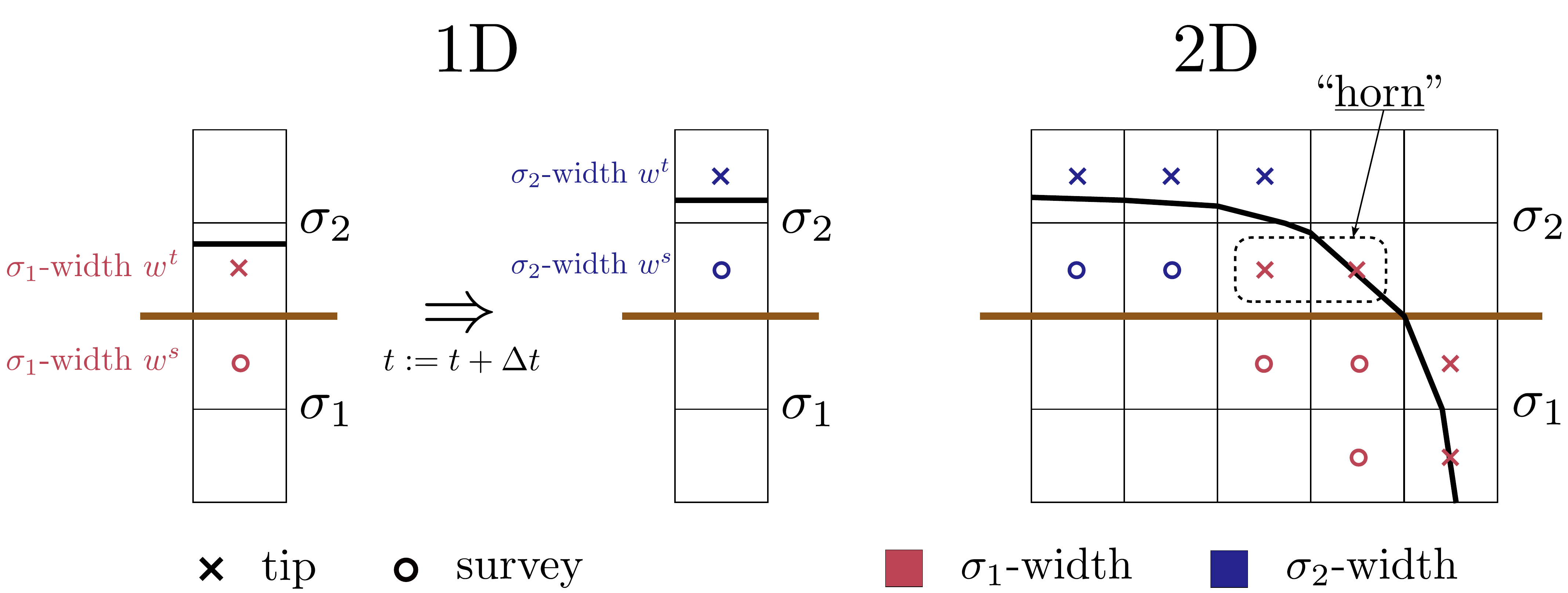}
            \caption{Schematics of the ``horns'' problem.}
            \label{img:horns_problem:schematic}
        \end{subfigure}
        \caption{The layer crossing or the ``horns'' problem for the case of strong stress barriers.}
        \label{img:horns_problem_illustration}
    \end{figure}

    Despite the aforementioned approach accounts for multiple physical effects, it may lead to significant errors in situations when the fracture front is crossing a stress layer since the effects of the layer are not included. The propagation condition is effectively enforced over two numerical elements, from a survey to a tip element. If a geologic layer is located between these two elements, its effect is completely ignored. To illustrate this phenomenon, consider an example with three stress layers where the outer layers have much higher confining stress than the middle layer. This corresponds to the situation of strong stress barriers and results in a nearly constant height fracture. The results of the calculations for such input parameters are shown in Figure~\ref{img:horns_problem:calculation}, where only half of a planar fracture is shown due to symmetry. The color filling shows the fracture aperture, while the yellow line outlines the fracture front. As can be seen from the figure, there is a numerical artifact located in the vicinity of the transition from the horizontal to vertical front orientation. There are tip elements next to the boundaries of the layers, whose width is significantly larger than the width of the neighboring elements located in the same layer. Further, we call such elements ``horns''. These ``horns'' also lead to the fracture front being outside the central layer by approximately one element above and below, thus leading to an overestimation of the fracture height.
    
    As was mentioned before, the ``horn'' or the layer crossing problem occurs because the asymptotic solution~\eqref{universal_asymptotic} does not account for the difference of the confining stress between the survey and the tip elements. To better understand the underlying mechanisms behind this phenomenon, the relevant schematic is depicted in Figure~\ref{img:horns_problem:schematic}. Here we have two stress layers, it is assumed that $\sigma_2 > \sigma_1$, and we focus exclusively on the vertical propagation. First, the survey element is located in the lower stress zone $\sigma_1$. The distance to the fracture front from this survey element is determined using equation~\eqref{universal_asymptotic}. But since the asymptotic solution~\eqref{universal_asymptotic} does not account for the presence of a stress layer, the distance to the fracture front $s$ is overestimated, which in turn leads to the overestimated width of the tip element $w^t$ (see $\sigma_1$-width $w^t$). Note that this tip element is located in the higher stress zone $\sigma_2$, and yet it does not ``feel'' it. Then for the next time step, let the fracture propagate so that the tip element becomes the new survey and the one above it becomes the new tip element. Now the survey element is affected by the high stress $\sigma_2$ and, consequently, its width becomes small (see $\sigma_2$-width $w^s$). This mechanism is responsible for the error in determining the fracture front location, and the error magnitude is on the order of element size. The right panel in Figure~\ref{img:horns_problem:schematic} shows this problem in a two-dimensional setting. The region containing the rotating fracture front has elements with an overestimated width (the so-called ``horn''). In relation to the previous one-dimensional example, this corresponds to the situation immediately before and promptly after the survey element jumped over the layer boundary. These schematics also closely resemble the situation observed in the actual numerical simulation shown in Figure~\ref{img:horns_problem:calculation}.

    To address the ``horns'' problem, it is necessary to consider a more general formulation for the problem of a semi-infinite hydraulic fracture. In particular, we need to consider the effect of multiple stress layers to overcome this issue. In the next section, we describe three various approaches to incorporate the effect of stress layers. These approaches differ in computational complexity, the complexity of implementation, and the accuracy of the approximation.
    
\section{Tip asymptotic solution with stress layers}\label{sec:tip_asymptotic_solutions}
    
    \subsection{Numerical solution using integral equation}
        The first approach represents the most accurate method and involves solving the problem of a steadily propagating semi-infinite hydraulic fracture with the presence of stress layers. In contrast to Section~\ref{sec:semi_infinite_fracture}, we further assume that the far-field confining stress $\sigma_0$ has a piece-wise constant spatial variation
        \begin{equation}
            \sigma_0(s) = \sigma_1 (1 - \mathcal{H}(s - s_1)) + \sum_{i=2}^{n} \sigma_i (\mathcal{H}(s - s_{i - 1}) - \mathcal{H}(s - s_{i})) + \sigma_{n + 1} \mathcal{H}(s - s_n),
        \end{equation}
        where $\mathcal{H}$ denotes Heaviside step-function, $s_j$ is the distance from the fracture tip to $j$th stress layer, and $\sigma_j$ denotes the magnitude of the stress in the layer located between $s_{i - 1}$ and $s_{i}$ for $i = 2, \ldots, n$, while the first and the last layers are considered to be semi-infinite. In this case, the non-singular integral formulation~\eqref{semi_infinite_integral_equation} can be extended to consider the stress layers in the fracture as~\cite{Dontsov_Homogenization_2017}
        \begin{equation}\label{integral_equation}
            \begin{aligned}
                w(s) = \frac{\Kprime}{\Eprime} s^{1/2} - &\frac{4}{\pi\Eprime} \int_{0}^{\infty} F(s, x')\frac{\muprime}{w(x')^2} \left[V + 2\Cprime V^{1/2}\frac{{x'}^{1/2}}{w(x')}\right] \mathrm{d}x' \\ - &\frac{4}{\pi\Eprime} \sum_{j=1}^{n} \Delta\sigma_j F(s, s_j),
            \end{aligned}
        \end{equation}
        where $\Delta\sigma_j = \sigma_{j} - \sigma_{j+1}$ is the amplitude of the $j$th stress change. The above formulation automatically satisfies the fracture propagation when $s\rightarrow 0$.

        Using the scaling~\eqref{semi_infinite_scaling} defined for a semi-infinite fracture and introducing the scaled amplitude of the stress barrier $\Delta\Sigma_j = \Delta\sigma_j l^{1/2} / \Kprime$, equation~\eqref{integral_equation} can be further reduced to
        \begin{equation}\label{integral_equation_dimless}
            \tildeW(\tildeS) = 1 + \frac{8}{\pi} \int_{0}^{\infty} G\left(\frac{\tilde{x}'}{\tildeS}\right) \left[\frac{1}{\tildeW(\tilde{x}')^2} + \frac{\chi}{\tildeW(\tilde{x}')^3}\right] \dint\tilde{x}' + \frac{4}{\pi} \sum_{j=1}^{n} \Delta\Sigma_j \tildeS_j G\left(\frac{\tildeS_j}{\tildeS}\right).
        \end{equation}
        In this scaling, the fracture propagation condition reduces simply to $\tildeW(0) = 1$.

        In order to obtain a numerical solution of~\eqref{integral_equation_dimless}, the integral is approximated using Simpson's rule and the resulting system of nonlinear algebraic equations is solved using Newton's method. The infinite upper integration limit is replaced by a sufficiently large finite number. The grid points used for the integral approximation are distributed uniformly on a logarithmic scale. The solution for the last several grid points is ignored to avoid additional numerical errors associated with the finite integration limit.
        
        While the solution of the integral equation~\eqref{integral_equation} can always be obtained numerically, it is still desirable to have a computationally faster method since the asymptotic solution is used many times within the fracture propagation algorithm. For instance, such an approach was used in~\cite{Dontsov_Homogenization_2017} for a plane strain fracture, for which only two tip elements are needed. Even in that case, the use of the integral equation significantly reduced the computational performance of the code. Unfortunately, it is not possible to obtain an approximate solution of~\eqref{integral_equation} using a similar approach as was used to obtain~\eqref{universal_asymptotic}. Therefore, we are going to explore other alternatives.
    
    \subsection{Toughness corrected asymptote}
        The second approach aims to be the most computationally efficient and involves using the universal asymptotic solution~\eqref{universal_asymptotic} with the corrected fracture toughness. Let us consider the simplified formulation of equation~\eqref{integral_equation} for the case of toughness-dominated propagation, i.e. excluding the terms responsible for viscous dissipation and fluid leak-off
        \begin{equation}\label{effective_width}
            w_{\mathrm{eff}}(s) = \frac{\Kprime}{\Eprime} s^{1/2} - \frac{4}{\pi\Eprime} \sum_{j=1}^{n} \Delta\sigma_j F(s, s_j).
        \end{equation}
        Using the above expression, we define effective toughness $\Kprime_{\mathrm{eff}}(s)$ as
        \begin{equation}\label{effective_toughness}
            \Kprime_{\mathrm{eff}}(s) = \frac{w_{\mathrm{eff}}(s) \Eprime}{s^{1/2}}.
        \end{equation}
        This effective toughness accounts for the effect of stress layers by changing the apparent toughness based on the value of the stress barrier and the distance to it. Equations~\eqref{universal_asymptotic} and~\eqref{effective_toughness} can be combined to yield the asymptotic solution that approximately accounts for layers with different confining stresses
        \begin{equation}\label{toughness_corrected_solution}
            \frac{s^{2} V \mu'}{E'w^{3}} = g_\delta \left(\frac{\Kprime_{\mathrm{eff}}(s) s^{1/2}}{E' w}, \frac{2s^{1/2} C'}{w V^{1/2}} \right).
        \end{equation}
        Typically, this equation would be solved using Newton's method. But Newton's method may not converge because the nonlinear function may be non-monotonic and non-convex in the vicinity of the root, especially when multiple layers are present. Therefore, it is necessary to choose an initial guess that is sufficiently close  to the unknown solution to ensure the convergence of Newton's method. Alternatively, it is possible to use the damped Newton method \cite{Ortega_Iterative_Methods_1970} or the Levenberg-Marquardt algorithm \cite{More_Levenberg_1978, Fan_Modified_Levenberg_Marquardt_2012}, which allows overcoming the local convergence of Newton's method.

        To avoid having negative effective fracture toughness $\Kprime_{\mathrm{eff}}(s)$, it is necessary to introduce a restriction on the magnitude of $\Delta\sigma_j$ as follows
        \begin{equation}\label{delta_sigma_restriction}
            \sum_{j=1}^{n} \Delta\sigma_j \frac{F(s, s_j)}{s^{1/2}} < \frac{\pi \Kprime}{4}.
        \end{equation}
        As can be seen from~\eqref{delta_sigma_restriction}, the restriction on $\Delta\sigma_j$ depends on the distance to the fracture front $s$. Given the fact that $F(s, s_j) \rightarrow -4s^{1/2}s_j^{1/2}$ as $s \rightarrow \infty$, the condition~\eqref{delta_sigma_restriction} for $s \gg s_j$ can be estimated as
        \begin{equation}\label{delta_sigma_restriction_large_x}
            \sum_{j=1}^{n} \Delta\sigma_j s_j^{1/2} > -\frac{\pi \Kprime}{16}.
        \end{equation}
        Such a restriction or the truncation of the stress layers (stress drops) introduces an error during layer crossing. At the same time, stress barriers (rather than stress drops) are arguably more important for fracture propagation, and they are captured more accurately.

        The computational complexity of the toughness-corrected asymptotic approach~\eqref{toughness_corrected_solution} is very similar to that of using a standard asymptote~\eqref{universal_asymptotic}. Yet, this approach allows us to account for the effect of stress layers. The method is precise only for toughness-dominated propagation and becomes approximate for other cases. This approach is somewhat similar to that used in~\cite{Dontsov_MuLTipEl_2022, Dontsov_MuLTipEl_SPE_2022}.
        
        Next, we are going to examine yet another approach, which lies in the middle between solving the integral equation and using the toughness correction. Namely, it aims to be less computationally intensive compared to the integral equation and yet more accurate than the simple toughness correction.
        
    \subsection{ODE approximation}
        The third and last approach is based on the ordinary differential equation (ODE) approximation of the original non-singular integral equation for the problem. Let us differentiate the dimensionless integral formulation~\eqref{integral_equation_dimless} and apply approximation $\tildeW \approx \tildeS^\delta$ similar to~\cite{Dontsov_Peirce_universal_asymptotic_2015} to obtain the ordinary differential equation
        \begin{equation}\label{ode_approx_singular}
            \opd{\tildeW}{\tildeS} = \frac{\beta_m^3}{3 \tildeW^2} + \frac{\chi \beta_{\tilde{m}}^4}{4 \tildeW^3} - \frac{4}{\pi} \sum_{j=1}^{n} \Delta\Sigma_j \left(\frac{\tildeS_j}{\tildeS}\right)^2 G'\left(\frac{\tildeS_j}{\tildeS}\right), \qquad \tildeW(0) = 1,
        \end{equation}
        where the kernel's derivative is
        \begin{equation}
            G'(t) = \frac{1}{t}\left(-\frac{1 + t^2}{t}\ln\left|\frac{1 + t}{1 - t}\right| + 2\right).
        \end{equation}
        
        Note that in the absence of the stress layers, the solution of equation~\eqref{ode_approx_singular} reduces to the zeroth-order approximate solution presented in~\cite{Dontsov_Peirce_universal_asymptotic_2015}. The zeroth-order solution has an accuracy of approximately 1.1 percent and is able to capture all vertex solutions precisely, while the $\delta$-correction~\cite{Dontsov_Peirce_universal_asymptotic_2015} increases the accuracy even further to 0.14 percent. In the presence of layers, the additional term on the right side of the equation~\eqref{ode_approx_singular} may cause a violation of the assumption $\tildeW \approx \tildeS^\delta$. Therefore, the use of $\delta$-correction for equation~\eqref{ode_approx_singular} is unlikely going to increase the accuracy of the approximation.
        
        It is worth noting that the kernel's derivative $G'(t)$ is singular at $t = 1$, which corresponds to $\tildeS = \tildeS_j$. To avoid this singularity, the following change of variable is performed
        \begin{equation}
            \widehat{w} = \tildeW - G_{\Sigma}(\tildeS), \quad G_{\Sigma}(\tildeS) = \frac{4}{\pi} \sum_{j=1}^{n} \Delta\Sigma_j \tildeS_j G\left(\frac{\tildeS_j}{\tildeS}\right)
        \end{equation}
        to yield the non-singular ODE formulation
        \begin{equation}\label{ode_approx_regular}
            \opd{\widehat{w}}{\tildeS} = \frac{\beta_m^3}{3(\widehat{w} + G_{\Sigma})^2} + \frac{\chi \beta_{\tilde{m}}^4}{4(\widehat{w} + G_{\Sigma})^3}.
        \end{equation}
        Since the integral kernel $G(t) \approx 4 / (3 t^2)$ for $t \gg 1$ then $G_{\Sigma}(0) = 0$ and the initial condition for the non-singular formulation~\eqref{ode_approx_regular} takes the following form
        \begin{equation}\label{ode_approx_regular_init_cond}
            \widehat{w}(0) = 1.
        \end{equation}
        The initial value problem~\eqref{ode_approx_regular} and~\eqref{ode_approx_regular_init_cond} is effectively solved numerically using explicit Dormand-Prince method~\cite{Dormand_Embedded_RK_methods_1980} with adaptive step size control~\cite{Macdonald_Adaptive_Step_RK_2001}.

        In the absence of the first and second terms on the right side of the equation~\eqref{ode_approx_singular}, the solution of the initial value problem~\eqref{ode_approx_singular} coincides with the solution~\eqref{effective_width}, which is used to calculate the effective fracture toughness~\eqref{effective_toughness}. The first and second terms on the right side of the equation~\eqref{ode_approx_singular} allow us to capture the influence of stress layers not only in the toughness-dominated regime but also in other regimes. In addition, there is no need to impose restrictions on the magnitude of $\Delta\Sigma_j$ to solve equation~\eqref{ode_approx_singular}. Thus, the ODE approximation is expected to be more accurate than the toughness-corrected approach. 
        
        To obtain the asymptotic width $\tildeW$ for a given $\tildeS$ using the ODE approximation, one needs to integrate the initial value problem~\eqref{ode_approx_regular},~\eqref{ode_approx_regular_init_cond} for the $[0, \tildeS]$ interval, which is done using the explicit Dormand-Prince method. In contrast, when using the integral equation, it is required to solve a coupled system of nonlinear equations and interpolate the solution at the point $\tildeS$. Hence, the ODE approximation is expected to be less computationally expensive compared to the integral equation approach.

\section{Comparison of the different approaches to include layers}\label{sec:comparison_of_approaches}
    In this section, we compare different approaches to calculate the tip asymptotic solution that accounts for the effect of stress layers. The purpose is to evaluate both accuracy and computational performance. The source code used in our analysis is available online at \href{https://github.com/alexander-valov/StressCorrectedAsymptote}{https://github.com/alexander-valov/StressCorrectedAsymptote}.
    
    To perform a comparison of the asymptotic solutions in the presence of stress layers, we consider the cases with stress jump $\Delta\Sigma_j > 0$ and stress drop $\Delta\Sigma_j < 0$. We also examine the accuracy of the solution for the so-called $k$-regime (or toughness-dominated propagation) and $m$-regime (or viscosity-dominated propagation). In particular, the regime is defined by considering a situation without the stress layer and is evaluated at the location of the stress layer $\tildeS_j$, see Figure~\ref{img:regimes_map}. In this section, we consider only the moderate level of leak-off, namely $\chi = 1$. To distinguish a weak and strong amplitude of the stress barrier for different regimes, let us introduce the relative influence of the stress barrier as
    \begin{equation}
        \Delta S_j = \frac{8 \Delta\Sigma_j \tildeS_j}{\pi \tildeW_0(\tildeS_j)},
    \end{equation}
    where $\tildeW_0(\tildeS)$ is the fracture width corresponding to the solution in a homogeneous formation. The latter equation represents the ratio of the impact of the $j$th stress layer $(4 / \pi) \Delta\Sigma_j \tildeS_j G(\tildeS_j / \tildeS)$ evaluated at the location of the stress
    layer $\tildeS = \tildeS_j$ to the fracture width $\tildeW_0(\tildeS)$ that ignores the stress layers. Thus, the amplitude of the stress barrier is considered weak if the relative influence of the stress barrier $|\Delta S_j| < 1 / 2$, while the strong amplitude of the stress barrier is defined as $|\Delta S_j| > 3 / 2$. The relative influence of the stress barrier is kept constant for the $k$ and $m$ regimes in calculations for a fair comparison.

\begin{figure}
        \begin{subfigure}{0.49\textwidth}
            \centering
            \includegraphics[width=1.0\linewidth]{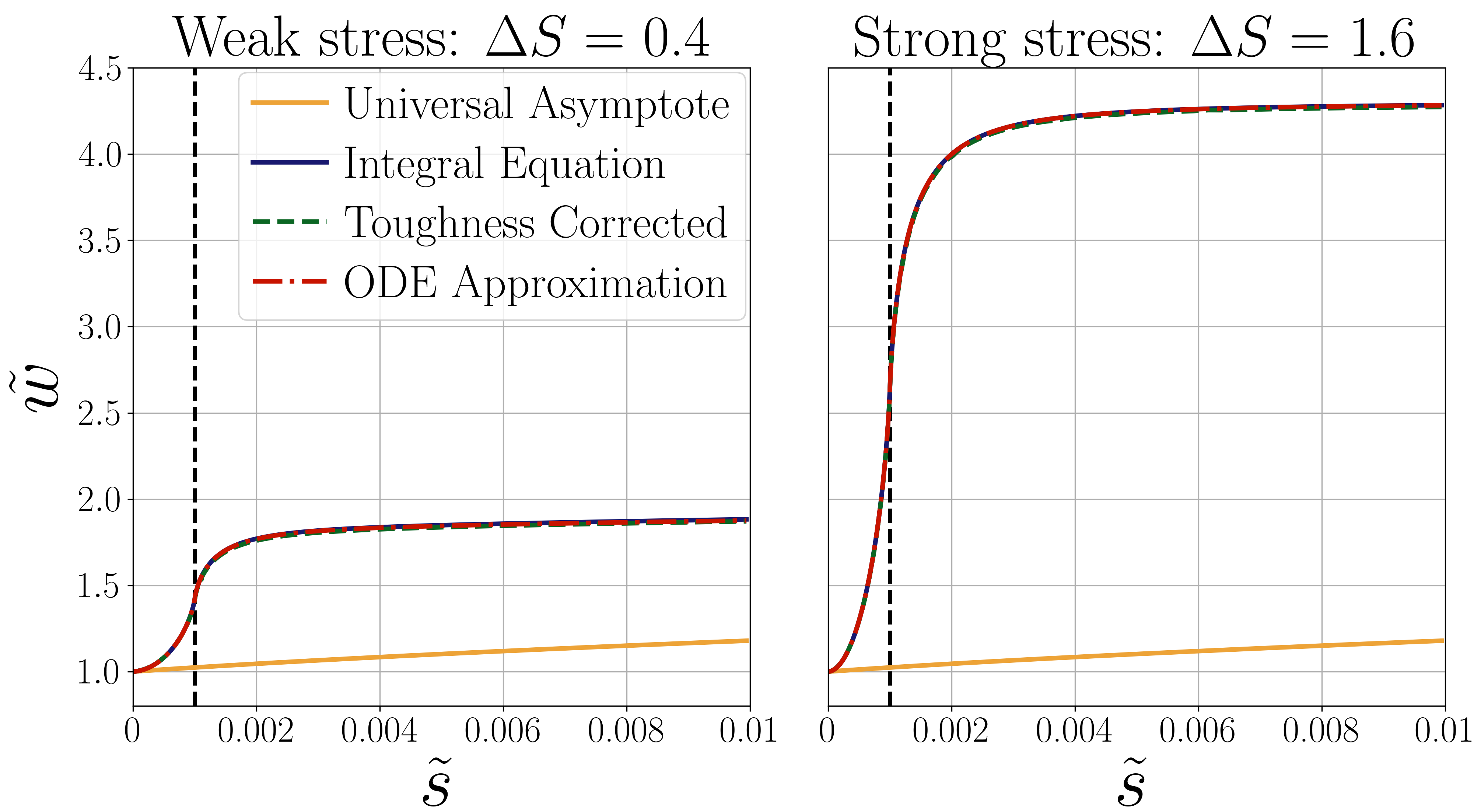}
            \caption{Stress jump, $k$-regime}
            \label{img:width_comparison_single_layer:jump_k}
        \end{subfigure}
        \begin{subfigure}{0.49\textwidth}
            \centering
            \includegraphics[width=1.0\linewidth]{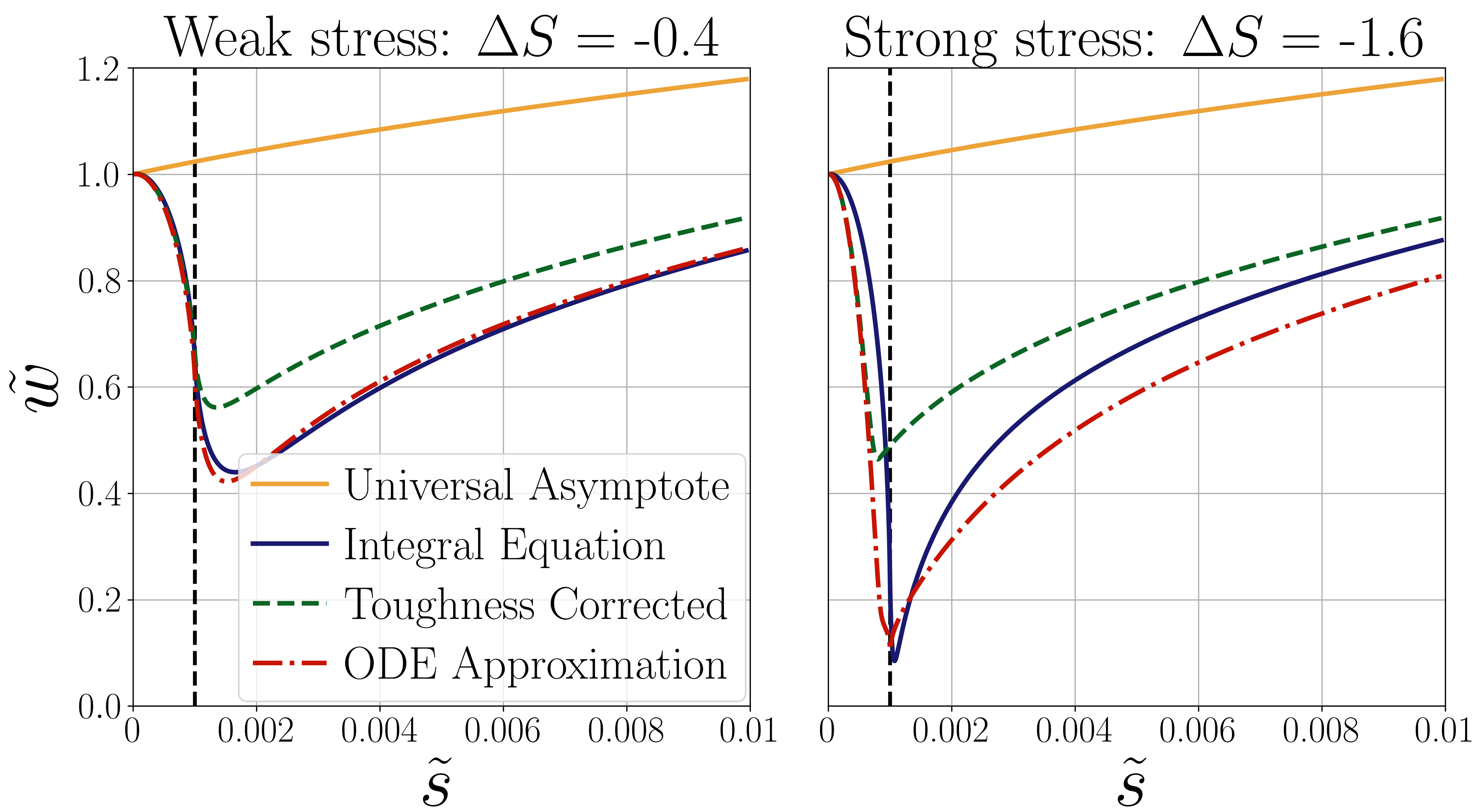}
            \caption{Stress drop, $k$-regime}
            \label{img:width_comparison_single_layer:drop_k}
        \end{subfigure}
        \\
        \begin{subfigure}{0.49\textwidth}
            \centering
            \includegraphics[width=1.0\linewidth]{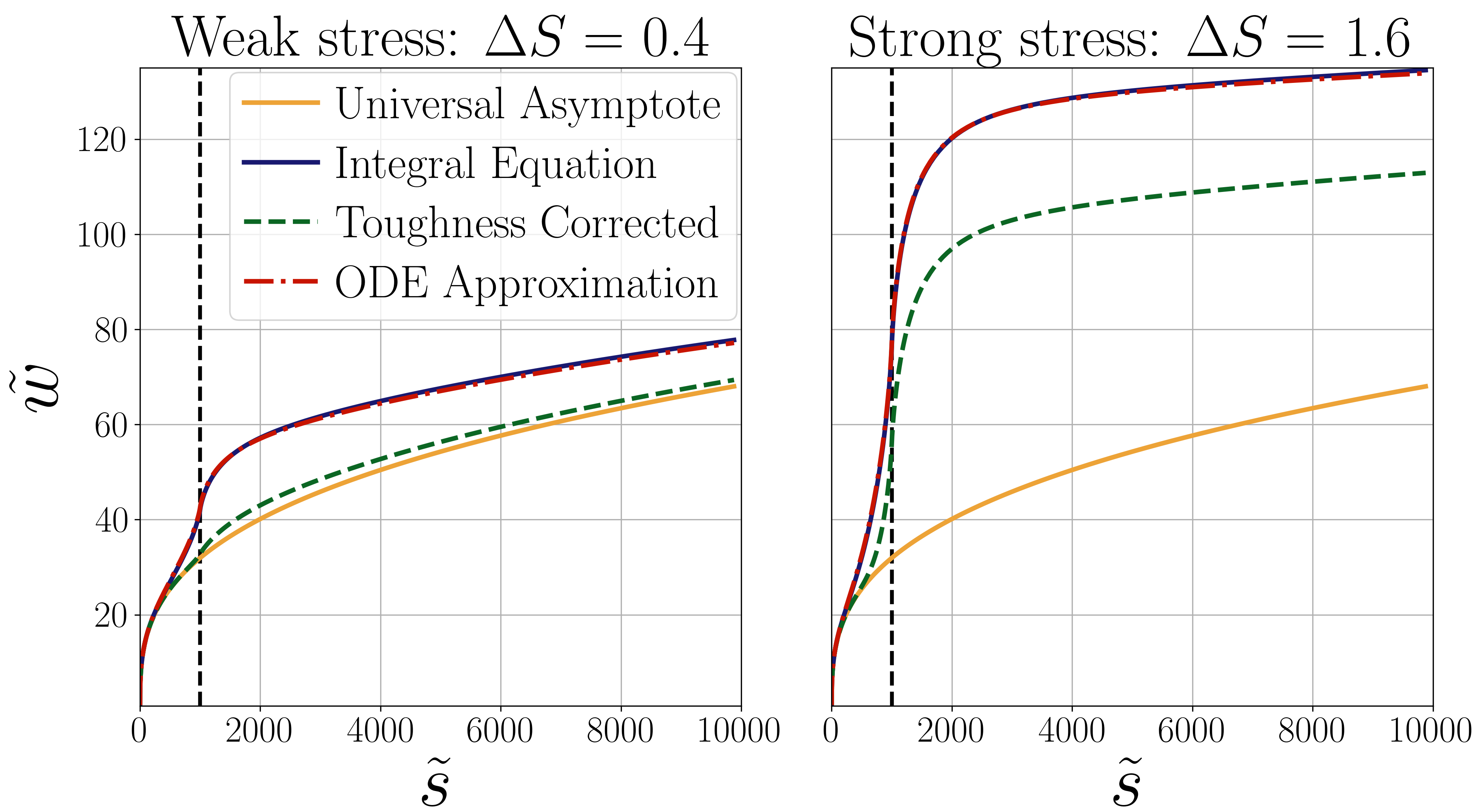}
            \caption{Stress jump, $m$-regime}
            \label{img:width_comparison_single_layer:jump_m}
        \end{subfigure}
        \begin{subfigure}{0.49\textwidth}
            \centering
            \includegraphics[width=1.0\linewidth]{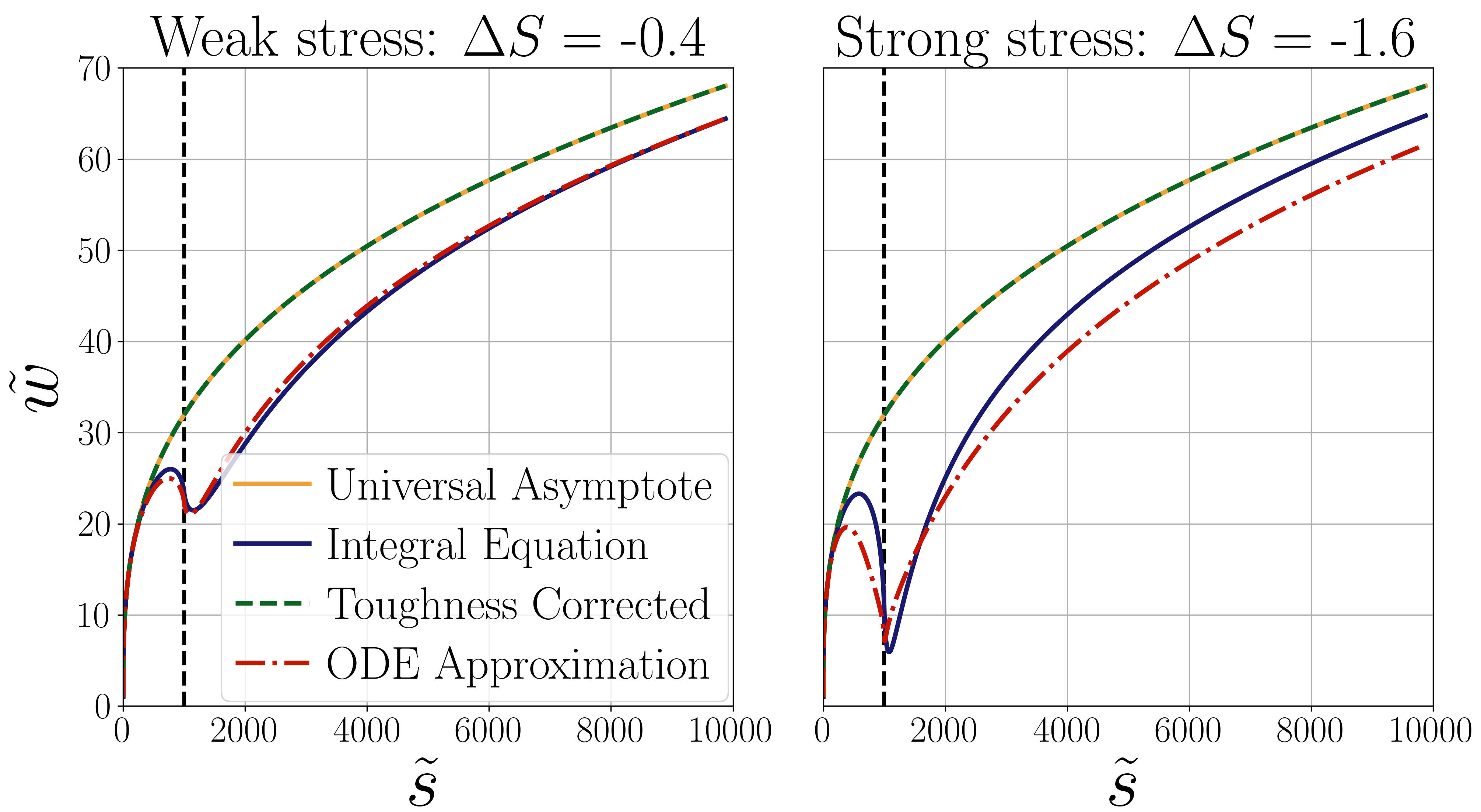}
            \caption{Stress drop, $m$-regime}
            \label{img:width_comparison_single_layer:drop_m}
        \end{subfigure}
        \caption{Comparison of the dimensionless width $\tildeW$ for the low and high stress jump and stress drop against dimensionless distance $\tildeS$ to the fracture front. The black dashed vertical lines show the boundary of layers. For the $k$-regime, the stress layer is located at $\tildeS_j = 0.001$, while for the $m$-regime at $\tildeS_j = 1000$.}
        \label{img:width_comparison_single_layer}
    \end{figure}

    Figure~\ref{img:width_comparison_single_layer} shows the solution for the dimensionless width $\tildeW$ versus scaled distance $\tildeS$ for a single stress jump and a single stress drop. All the approaches summarized in the previous section are included: numerical solution using integral equation (the solid blue lines), toughness corrected asymptote (the dashed green lines), and ODE approximation (the dash-dotted red lines). In addition, the universal asymptotic solution (the solid yellow line) is shown to demonstrate the contribution of stress to the fracture width. The left panel in each figure corresponds to the weak stress barrier, while the right one corresponds to the strong stress barrier.

    We start with the $k$-regime and observe the comparison of the asymptotic solutions for a stress layer located at $\tildeS_j = 0.001$ and other parameters corresponding to the $k$-regime. The  amplitude of the weak stress barrier corresponds to $|\Delta\Sigma_j| = 161$, while the strong barrier has $|\Delta\Sigma_j| = 643$. As can be seen from Figure~\ref{img:width_comparison_single_layer:jump_k}, there is no visual difference between the ODE approximation and the precise integral equation solution for the case of stress jump. The toughness-corrected approximation is also in good agreement with the solution of the integral equation. Moreover, the higher the stress jump is, the better the agreement becomes. Recall that the toughness correction is precise for the toughness-dominated propagation and therefore this result is expected. Negative values of the amplitude of the $j$th stress layer $\Delta\Sigma_j$ introduce the non-monotonic behavior to the solution, which makes it more challenging for the numerical algorithms to converge for large magnitudes of stress drops. Figure~\ref{img:width_comparison_single_layer:drop_k} shows the comparison of the scaled width $\tildeW$ for the negative values of $\Delta\Sigma_j$. For the case of a small stress drop, the ODE approximation matches closely with the integral equation solution. However, for the case of high magnitude of the stress drop, the ODE approximation matches the integral equation solution only qualitatively. The toughness-corrected solution noticeably underestimates the influence of the stress drop and considerably differs from the precise integral equation solution.

    The $m$-regime is considered next. It corresponds to the stress layer located at $\tildeS = 1000$. The amplitude of the weak stress barrier corresponds to $|\Delta\Sigma_j| = 0.005$, while the strong barrier has $|\Delta\Sigma_j| = 0.02$. As can be seen from Figure~\ref{img:width_comparison_single_layer:jump_m}, the ODE approximation and integral equation solution provide nearly identical results for the case of stress jump. But the toughness corrected approximation is much less accurate since the effect of viscosity plays a significant role in the $m$-regime. For the negative values of the amplitude of the $j$th stress layer $\Delta\Sigma_j$, see Figure~\ref{img:width_comparison_single_layer:drop_m}, when the relative stress drop is not so significant, the ODE approximation is in a good agreement with the solution of the integral equation. When the relative stress drop is increased, the ODE approximation becomes less accurate, similar to the toughness-dominated case. The toughness-corrected solution for both small and large magnitudes of the stress drop is practically insensitive to the stress layer and coincides with the universal asymptotic solution. This is expected since the solution does not accurately capture the effect of viscosity.
    
    It is important to note that the toughness corrected approximation differs noticeably from the integral equation solution for the $k$-regime and performs even poorer for the $m$-regime, especially for the stress drop due to the limitation $\Kprime_\mathrm{eff}(s) > 0$ on the effective toughness. According to~\eqref{delta_sigma_restriction}, for the case of a single layer, the following limitation on the amplitude of the stress barrier $\Delta\sigma$ is met
    \begin{equation}\label{delta_sigma_limitation}
        \Delta\sigma > \frac{\pi \Kprime}{4} \frac{s^{1/2}}{F(s, s_j)},
    \end{equation}
    where $F(s, s_j) < 0$ for all $s, s_j > 0$. For $s \gg s_j$, instead of the relation~\eqref{delta_sigma_limitation}, the following expression can be used to estimate the lower bound of $\Delta\sigma$
    \begin{equation}\label{delta_sigma_limitation_large_x}
        \Delta\sigma > -\frac{\pi \Kprime}{16 s_j^{1/2}}.
    \end{equation}
    Thus, the toughness-corrected asymptote allows the stress drop to be taken into account only for relatively small distances to the stress
    layer $\tildeS_j$.

    In order to estimate the approximation error let us introduce the relative error as
    \begin{equation}
        \varepsilon_2 = \frac{\norm{\tildeW_\mathrm{num} - \tildeW_\mathrm{int}}_2}{\norm{\tildeW_\mathrm{int}}_2}, \qquad \norm{\tildeW}_2 = \left(\int\limits_{0}^{\tildeS_\mathrm{max}} \tildeW^2 \dint\tildeS \right)^{1/2},
    \end{equation}
    where $\tildeW_\mathrm{int}$ is the integral equation solution, $\tildeW_\mathrm{num}$ is the toughness corrected solution or ODE approximation, $\tildeS_\mathrm{max} = 0.01$ for the $k$-regime, and $\tildeS_\mathrm{max} = 10^4$ for the $m$-regime.
    
    Table~\ref{tab:asymptotic_relative_error} shows the relative error of the toughness-corrected solution and the ODE approximation presented in Figure~\ref{img:width_comparison_single_layer}. The relative error of the toughness corrected solution for stress jump in the $k$-regime does not exceed 1 percent. However, for other cases, the relative error is significantly bigger and reaches 13-20 percent. The ODE approximation shows high accuracy for the case of stress jump for both $k$-regime and $m$-regime, and the relative error does not exceed 1 percent. For the case of moderate stress drop, the approximation error stays within 2 percent. The relative error for the high stress drop increases to 7-15 percent. But it is worth noting that despite having a noticeable error, the ODE approximation is still able to qualitatively describe the integral equation solution.

    \begin{table}
        \setlength\tabcolsep{0pt}

        \begin{subtable}[t]{1.0\textwidth}
            \centering
            \begin{tabular*}{\textwidth}{@{\extracolsep{\fill}}lccccccc}
                \toprule
                & \multicolumn{2}{c}{Stress jump} & \multicolumn{2}{c}{Stress drop} \\
                \cline{2-3} \cline{4-5}
                $\Delta S$                        & 0.4       & 1.6       & -0.4       & -1.6      \\
                \midrule
                Toughness $\varepsilon_2$         & 0.63 \%   & 0.26 \%   & 13.64 \%   & 19.25  \% \\
                ODE $\varepsilon_2$               & 0.24 \%   & 0.02 \%   & 1.72  \%   & 15.52  \% \\
                \bottomrule
            \end{tabular*}
            \caption{Relative error for the $k$-regime}
        \end{subtable}

        \bigskip

        \begin{subtable}[h]{1.0\textwidth}
            \centering
            \begin{tabular*}{\textwidth}{@{\extracolsep{\fill}}lccccccc}
                \toprule
                & \multicolumn{2}{c}{Stress jump} & \multicolumn{2}{c}{Stress drop} \\
                \cline{2-3} \cline{4-5}
                $\Delta S$                        & 0.4        & 1.6        & -0.4       & -1.6      \\
                \midrule
                Toughness $\varepsilon_2$         & 16.59 \%   & 17.70 \%   & 14.61 \%   & 20.03 \%  \\
                ODE $\varepsilon_2$               & 0.77  \%   & 0.34  \%   & 1.27  \%   & 7.77  \%  \\
                \bottomrule
            \end{tabular*}
            \caption{Relative error for the $m$-regime}
        \end{subtable}
        \caption{The approximation error for the toughness corrected solution and ODE approximation relative to the integral equation solution for the $k$-regime and $m$-regime. }
        \label{tab:asymptotic_relative_error}
    \end{table}

\begin{figure}
        \begin{subfigure}{1.0\textwidth}
            \centering
            \includegraphics[width=1.0\linewidth]{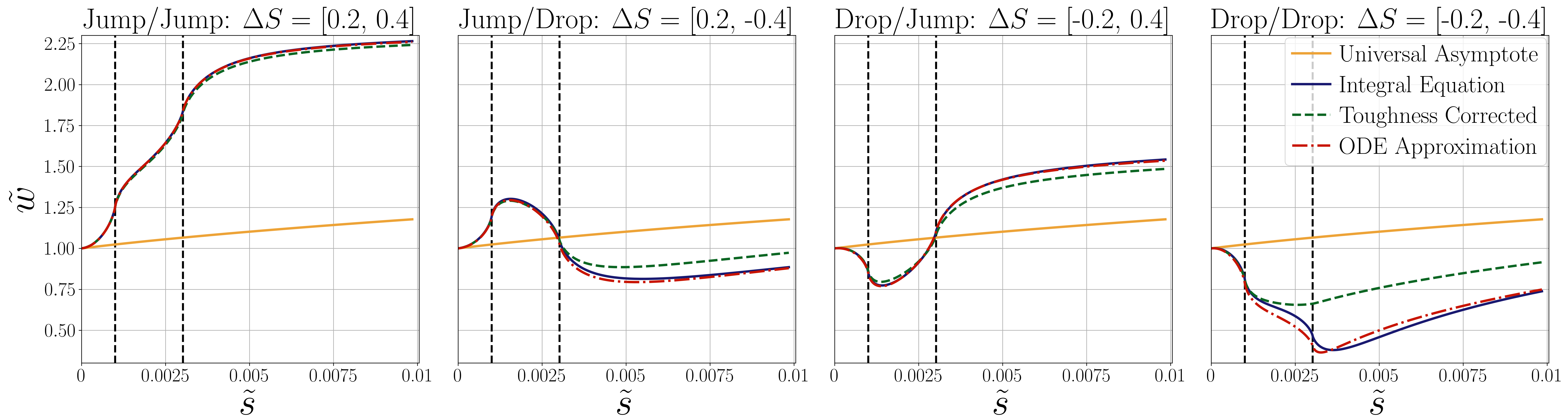}
            \caption{$k$-regime, the stress layers are located at $\tildeS_1 = 0.001$ and $\tildeS_2 = 0.003$}
            \label{img:width_comparison_two_layers:k}
        \end{subfigure}
        \\
        \begin{subfigure}{1.0\textwidth}
            \centering
            \includegraphics[width=1.0\linewidth]{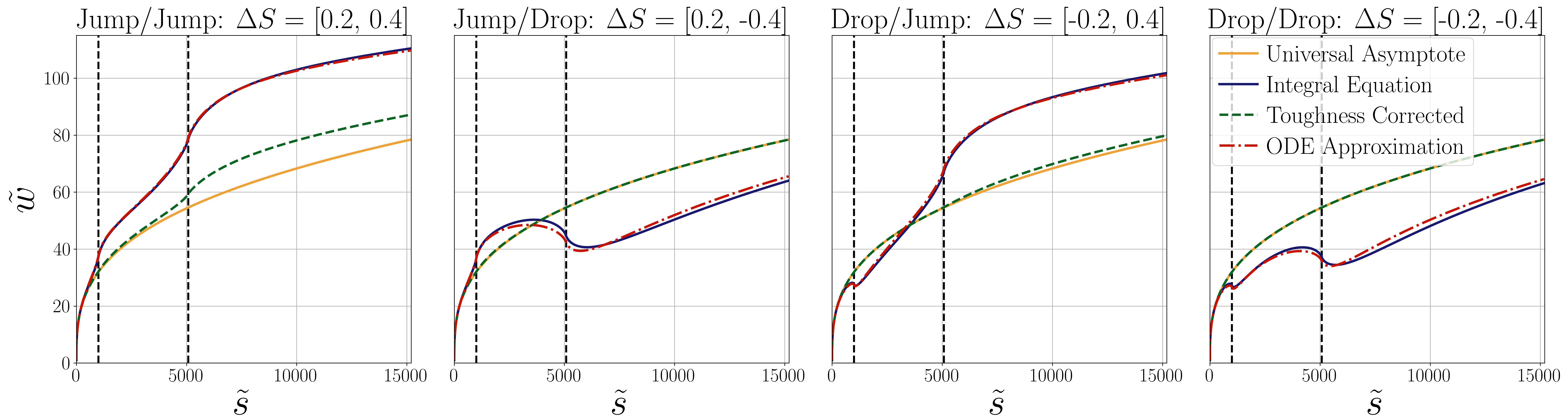}
            \caption{$m$-regime, the stress layers are located at $\tildeS_1 = 1000$ and $\tildeS_2 = 5080$}
            \label{img:width_comparison_two_layers:m}
        \end{subfigure}
        \caption{Comparison of the dimensionless width $\tildeW$ in the case of two stress layers against dimensionless distance $\tildeS$ to the fracture front. The black dashed vertical lines show the boundary of layers. }
        \label{img:width_comparison_two_layers}
    \end{figure}

    Figures~\ref{img:width_comparison_two_layers:k}--\ref{img:width_comparison_two_layers:m} show the comparison of the asymptotic solutions for the case of two layers: jump/jump, jump/drop, drop/jump, and drop/drop. For the case of two stress layers, in addition to keeping the relative influence of the stress barriers $\Delta S_j$ constant for each layer, the stress ratio $\Delta\Sigma_1 / \Delta\Sigma_2$ is also kept constant. For the $k$-regime, the amplitudes of the dimensionless stress barriers are equal to $|\Delta\Sigma_1| = 80$ and $|\Delta\Sigma_2| = 54$, while for the $m$-regime they are $|\Delta\Sigma_1| = 0.0025$ and $|\Delta\Sigma_2| = 0.0017$. As can be seen from Figure~\ref{img:width_comparison_two_layers:k}, in the $k$-regime for the jump/jump case, both the toughness corrected solution and the ODE approximation match the integral equation solution closely. For the case of the jump/drop, toughness corrected approximation follows the integral equation solution near the stress jump. However, there is a difference after the stress drop due to the limitation $\Kprime_\mathrm{eff} > 0$. For the case of drop/jump, the toughness-corrected solution slightly underestimates the magnitude of the stress jump at the second layer due to the stress drop at the first layer. Finally, for the case of drop/drop, the toughness-corrected solution ignores the second stress layer, while the ODE approximation qualitatively coincides with the solution of the integral equation. 
    
    As shown in Figure~\ref{img:width_comparison_two_layers:m}, for the $m$-regime the ODE approximation is in good agreement with the integral equation solution for any combination of layers. At the same time, the toughness-corrected asymptote almost entirely ignores the presence of stress layers except for the jump/jump case.

    To compare the computational performance of the approaches described above, we perform calculations for both $k$-regime and $m$-regime for different values of $\Delta\Sigma_j$. For the numerical solution of the integral equation and the toughness corrected approximation, the maximum of the vertex solutions~\eqref{vertex_solutions_homogeneous}, i.e. $\max(\tildeW_{k}, \tildeW_{\tilde{m}}, \tildeW_{m})$ is used as an initial approximation for Newton's method. For the numerical solution of the ODE approximation, we use the explicit Dormand-Prince method with adaptive step size control. The average number of nodes for the prescribed tolerances is equal to 9 for $k$-regime and 26 for $m$-regime. For the numerical solution of the integral equation, we use the fixed grid covering the interval $0 \leqslant \tildeS \leqslant 10^{18}$ with 1500 nodes distributed on a logarithmic scale and condensed in the vicinity of the stress layers. 
    
    Table~\ref{tab:asymptotic_performance_comparison} shows the comparison of the run times. The most computationally efficient approach of the presented solutions is the toughness-corrected approach. The ODE approximation has a calculation efficiency that is lower but has a similar order of magnitude. Note that the calculation time for the ODE approximation is increased in the $m$-regime compared to the $k$-regime. This performance reduction is a direct consequence of the computational domain extension (i.e. the solution is calculated for larger values of the scaled distance $\tildeS$). The numerical solution of the integral equation requires approximately two orders of magnitude more calculation time than the computation for the other approximations. This is a significant difference that makes the integral equation approach less suitable for the implementation of fracture front tracking.

    \begin{table}
        \setlength\tabcolsep{0pt}
        \begin{subtable}[t]{\textwidth}
            \centering
            \begin{tabular*}{\textwidth}{@{\extracolsep{\fill}}lccccccc}
                \toprule
                $\Delta S$    & -1 & -0.5 & -0.25 & 0 & 0.25 & 0.5 & 1 \\
                \midrule
                Integral, ms   &  1078.04 & 302.70 &  193.28 &  194.75 &  194.26 &  229.63 &  231.48 \\
                Toughness, ms  &  1.61    & 1.60 & 1.50 & 1.26 & 1.67 & 2.11 & 2.86 \\
                ODE, ms        &  11.35   & 4.68 &  2.32 &  0.47 &  1.97 &  2.23 &  2.01 \\
                \bottomrule
            \end{tabular*}
            \caption{Computation time for $\tildeS_j = 0.001$}
            \label{tab:asymptotic_performance_comparison:k}
        \end{subtable}

        \bigskip

        \begin{subtable}[h]{0.99\textwidth}
            \centering
            \begin{tabular*}{\textwidth}{@{\extracolsep{\fill}}lccccccc}
                \toprule
                $\Delta S$ & -1 & -0.5 & -0.25 & 0 & 0.25 & 0.5 & 1 \\
                \midrule
                Integral, ms  &  736.27 & 190.28 & 191.62 & 191.13 & 190.98 & 192.96 & 192.13 \\
                Toughness, ms  &  1.08 & 1.08 & 1.09 & 1.09 & 1.15 & 1.30 & 1.39 \\
                ODE, ms  &  11.88 &  8.85 &  8.03 &  6.35 &  7.30 &  7.29 &  7.52 \\
                \bottomrule
            \end{tabular*}
            \caption{Computation time for $\tildeS_j = 1000$}
            \label{tab:asymptotic_performance_comparison:m}
        \end{subtable}
        \caption{Comparison of the computation time of the integral equation, toughness corrected asymptote, and ODE approximation for layer locations $\tildeS_j = 0.001$ and $\tildeS_j = 1000$. }
        \label{tab:asymptotic_performance_comparison}
    \end{table}

    Based on the results in this section, we summarize the advantages and disadvantages of each approach. The advantage of the numerical solution of the integral equation includes superior accuracy. While the primary disadvantage of this approach is the high computational cost, which makes it impractical to use this method for front tracking in a hydraulic fracturing simulator, such as ILSA. The advantages of the toughness-corrected approach include simplicity of implementation, satisfactory accuracy for the $k$-regime, and computational efficiency. The disadvantages of this approach include a high level of error for the parameters that are outside the $k$-regime and an almost complete inability to capture a stress drop for the $m$-regime. The ODE approximation represents the middle ground between the other two approaches. It adequately captures both stress jump and stress drop cases, as well as all two-layer cases. Its computational efficiency is lower, but it is still comparable to the toughness-corrected approach. The primary disadvantage includes a significant error in the case of high stress drop.

\section{Validity region of the asymptotic solutions for a finite fracture}\label{sec:validity_region}
\begin{figure}
        \begin{subfigure}{0.49\textwidth}
            \centering
            \includegraphics[width=1.0\linewidth]{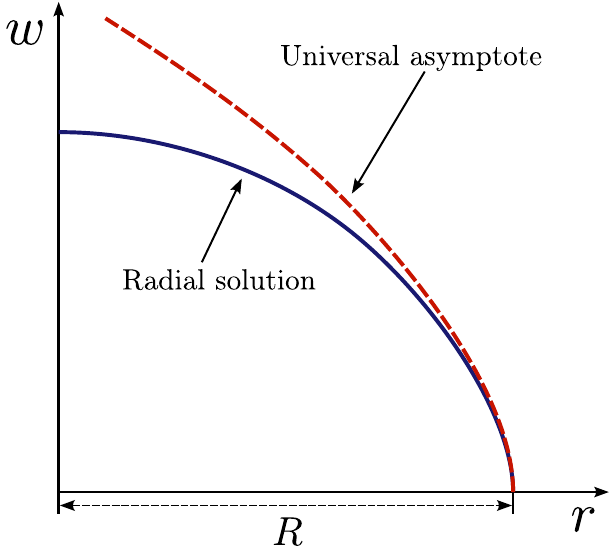}
            \caption{Radial fracture}
            \label{img:validity_region_scheme:radial}
        \end{subfigure}
        \begin{subfigure}{0.49\textwidth}
            \centering
            \includegraphics[width=1.0\linewidth]{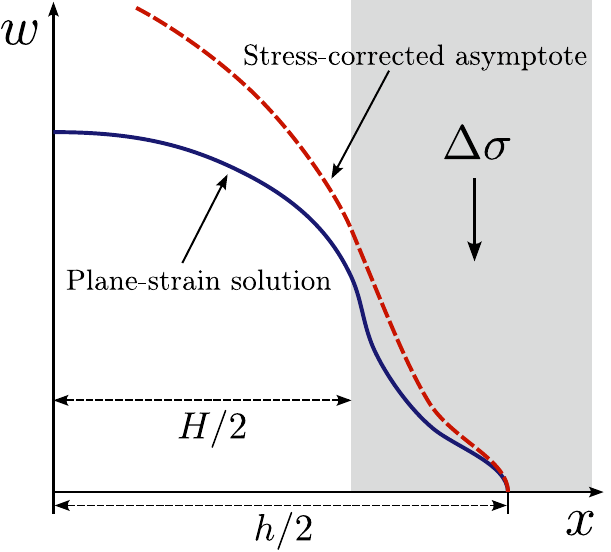}
            \caption{Plane-strain fracture}
            \label{img:validity_region_scheme:plane_strain}
        \end{subfigure}
        \caption{Schematics of a fracture width for the radial geometry compared to the universal asymptotic solution and the solution for a plane-strain fracture subject to two symmetric stress barriers compared to the stress-corrected asymptotic solution. Only half of a fracture is shown due to symmetry.}
        \label{img:validity_region_scheme}
    \end{figure}

    The addressed problem of a semi-infinite fracture with stress layers is applicable to the tip region of a finite fracture. As was mentioned in~\cite{Dontsov_Peirce_ILSA_2017}, the actual region of validity of the universal asymptotic solution~\eqref{universal_asymptotic} in a homogeneous formation occupies a few elements in the tip region, see Figure~\ref{img:validity_region_scheme:radial}. This fact makes it possible to use the universal asymptotic solution on a relatively coarse mesh without significant loss of accuracy. For the case of the layer crossing problem, the typical size of the ``horn'' is approximately one element, so the problem is especially critical on a coarse mesh. For this reason, it is necessary to investigate the accuracy of the tip asymptote with stress layers (or stress-corrected asymptote) on a coarse grid. Therefore, the purpose of this section is to quantify the size of the tip region, in which the asymptotic solution is still accurate.
    
    In order to determine the validity region for the asymptotic solutions, we consider a uniformly pressurized plane-strain hydraulic fracture, subject to two symmetric stress barriers~\cite{Adachi_Detournay_Classical_P3D_model_2010, Dontsov_Peirce_EP3D_2015}. The fracture height is $h$, the reservoir layer height is $H$, while the stress barriers have additional confining stress $\Delta\sigma$, as depicted in Figure~\ref{img:validity_region_scheme:plane_strain}. The use of the constant pressure assumption allows us to calculate the fracture width profile explicitly for a given fracture height $h$ as follows
    \begin{equation}\label{validity_finite_fracture_width}
        w_{\mathrm{ps}}(x) = \frac{2}{\Eprime}\sqrt{\frac{2}{\pi h}}K_\mathrm{Ic}\varphi + \frac{4\Delta\sigma}{\pi\Eprime}\left\{
            -x\ln{\left| \frac{H\varphi + 2x\psi}{H\varphi - 2x\psi} \right|} +
            \frac{H}{2}\ln{\left| \frac{\varphi + \psi}{\varphi - \psi} \right|}
        \right\},
    \end{equation}
    where $\varphi = \left(h^2 - 4 x^2\right)^{1/2}$, $\psi = \left(h^2 - H^2\right)^{1/2}$. The uniform pressure assumption implicitly implies that the fracture propagates in the toughness regime. Due to the symmetry of the problem, we consider only the right half of the fracture $x \geq 0$.

    In the case of a single stress barrier for the toughness regime, the integral equation~\eqref{integral_equation}, toughness-corrected solution~\eqref{toughness_corrected_solution}, and ODE approximation~\eqref{ode_approx_regular} all reduce to
    \begin{equation}\label{validity_tip_asymptote}
        w_{\mathrm{a}}(s) = \frac{\Kprime}{\Eprime} s^{1/2} - \frac{4}{\pi\Eprime} \Delta\sigma F(s, \xi),
    \end{equation}
    where $s = h / 2 - x$ is the distance to the fracture front and $\xi = (h - H) / 2$ is the distance from the stress barrier to the fracture tip.

    Before quantifying the validity region of the asymptotic solution, it is convenient to scale the problem as follows:
    \begin{equation}\label{validity_scaling}
        \widebar{w} = \frac{\Eprime w}{K_\mathrm{Ic} H^{1 / 2}}, \quad \widebar{h} = \frac{h}{H}, \quad \widebar{x} = \frac{x}{H}, \quad \widebar{s} = \frac{s}{H}.
    \end{equation}
    With these definitions, all the results depend on a single dimensionless parameter representing the amplitude of the stress barrier
    \begin{equation}\label{validity_scaled_stress}
        \Delta\widebar{\sigma} = \frac{\Delta\sigma H^{1/2}}{K_\mathrm{Ic}}.
    \end{equation}
    The scaled plane-strain fracture width profile and scaled asymptotic width take the following form
    \begin{gather}
        \widebar{w}_\mathrm{ps} = \sqrt{\frac{8}{\pi \widebar{h}}} \, \widebar{\varphi} + \frac{4 \Delta\widebar{\sigma}}{\pi} \left\{
            -\widebar{x}\ln{\left| \frac{\widebar{\varphi} + 2\widebar{x}\widebar{\psi}}{\widebar{\varphi} - 2\widebar{x}\widebar{\psi}} \right|} +
            \frac{1}{2}\ln{\left| \frac{\widebar{\varphi} + \widebar{\psi}}{\widebar{\varphi} - \widebar{\psi}} \right|}
        \right\}, \\
        \widebar{w}_\mathrm{a} = \sqrt{\frac{32}{\pi}} \, \widebar{s}^{1/2} - \frac{4 \Delta\widebar{\sigma}}{\pi} F(\widebar{s}, \widebar{\xi}),
    \end{gather}
    where $\widebar{\varphi} = \left(\widebar{h}^2 - 4 \widebar{x}^2\right)^{1/2}$, $\widebar{\psi} = \left(\widebar{h}^2 - 1\right)^{1/2}$, and $\widebar{\xi} = (\widebar{h} - 1) / 2$. We define the validity region $\rho_v $ of the stress-corrected asymptote as the ratio of the distance from the fracture tip at which the relative error $|(\widebar{w}_{\mathrm{ps}} - \widebar{w}_{\mathrm{a}}) / \widebar{w}_{\mathrm{ps}}|$ is below 5 percent to the fracture half height $\widebar{h} / 2$. For instance, the validity region is 0.5 if the error of the tip asymptote is 5\% in the middle of the fracture.

\begin{figure}
        \centering
        \includegraphics[width=0.99\linewidth]{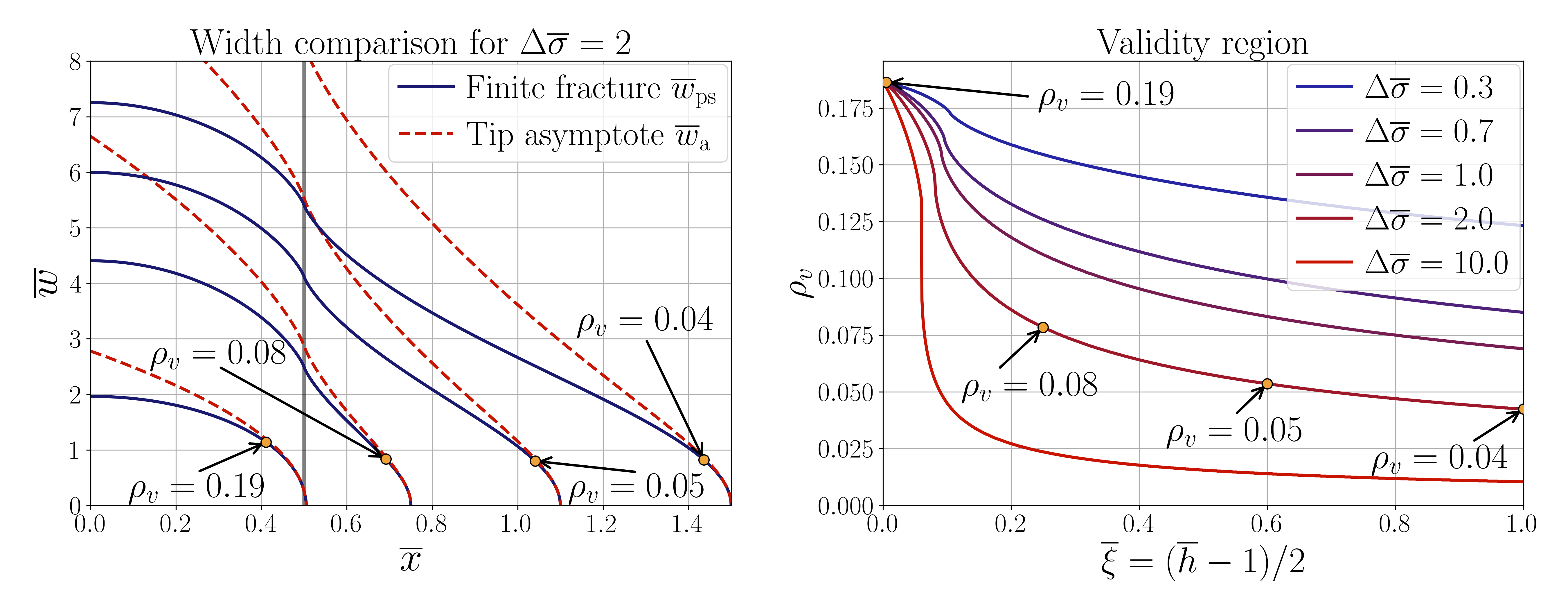}
        \caption{Comparison between the finite fracture solution~\eqref{validity_finite_fracture_width} and the tip asymptotic solution~\eqref{validity_tip_asymptote} for several values of fracture height (left). The yellow markers indicate the validity region. The right panel shows the variation of the validity region versus the distance from the fracture tip to the layer for different amplitudes of the stress barrier $\Delta\widebar{\sigma}$.}
        \label{img:validity_region}
    \end{figure}

    The left panel in Figure~\ref{img:validity_region} shows the comparison of the finite fracture solution~\eqref{validity_finite_fracture_width} and the tip asymptotic solution~\eqref{validity_tip_asymptote} for the fixed amplitude of the stress barrier $\Delta\widebar{\sigma} = 2$. As the fracture height increases, the error between the finite fracture solution $\widebar{w}_{\mathrm{ps}}$ and the tip asymptote $\widebar{w}_{\mathrm{a}}$ also increases, and the validity region $\rho_v$ decreases significantly from 19 percent for height $\widebar{h} = 1$ to 4 percent for height $\widebar{h} = 3$. The right panel in Figure~\ref{img:validity_region} depicts the region of validity $\rho_v$ for the different amplitudes of the stress barrier $\Delta\widebar{\sigma}$ and various fracture heights $\widebar{h}$. For small amplitudes of the stress barrier $\Delta\widebar{\sigma}$, the validity region changes slightly with increasing fracture height. While for large amplitudes of the stress barrier $\Delta\widebar{\sigma}$, the region of validity $\rho_v$ decreases drastically when the fracture tip moves away from the layer boundary.

    In order to overcome the drastic decrease of the validity region when the fracture height increases, we modify the asymptotic solution by introducing the stress relaxation factor $\lambda(\widebar{\xi})$ as follows
    \begin{equation}\label{validity_stress_relaxation_asymptote}
        \widebar{w}_{\mathrm{a}}^{\lambda}(\widebar{s}) = \sqrt{\frac{32}{\pi}} \, \widebar{s}^{1/2} - \frac{4 \lambda(\widebar{\xi}) \Delta\widebar{\sigma}}{\pi} F(\widebar{s}, \widebar{\xi}).
    \end{equation}
    The stress relaxation factor $\lambda$ is determined by minimizing the average difference between the finite fracture solution~\eqref{validity_finite_fracture_width} and the stress-corrected asymptote with the stress relaxation factor~\eqref{validity_stress_relaxation_asymptote} as
    \begin{equation}\label{lambda_minimization}
        \lambda = \argmin_{\lambda_\ast} \left\{ \int_{0}^{L} \Bigl( \widebar{w}_{\mathrm{ps}}(\widebar{h} / 2 - \widebar{s}) - \widebar{w}_{\mathrm{a}}^{\lambda_\ast}(\widebar{s}) \Bigr)^2 \dint \widebar{s} \right\}
    \end{equation}
    for different magnitudes of the fracture height $\widebar{h}$ and the amplitudes of the stress barrier $\Delta\widebar{\sigma}$. The upper limit $L$ of the integral in~\eqref{lambda_minimization} signifies the size of the tip region and varies in calculations to imitate the different mesh sizes.

    \begin{figure}
        \centering
        \includegraphics[width=0.99\linewidth]{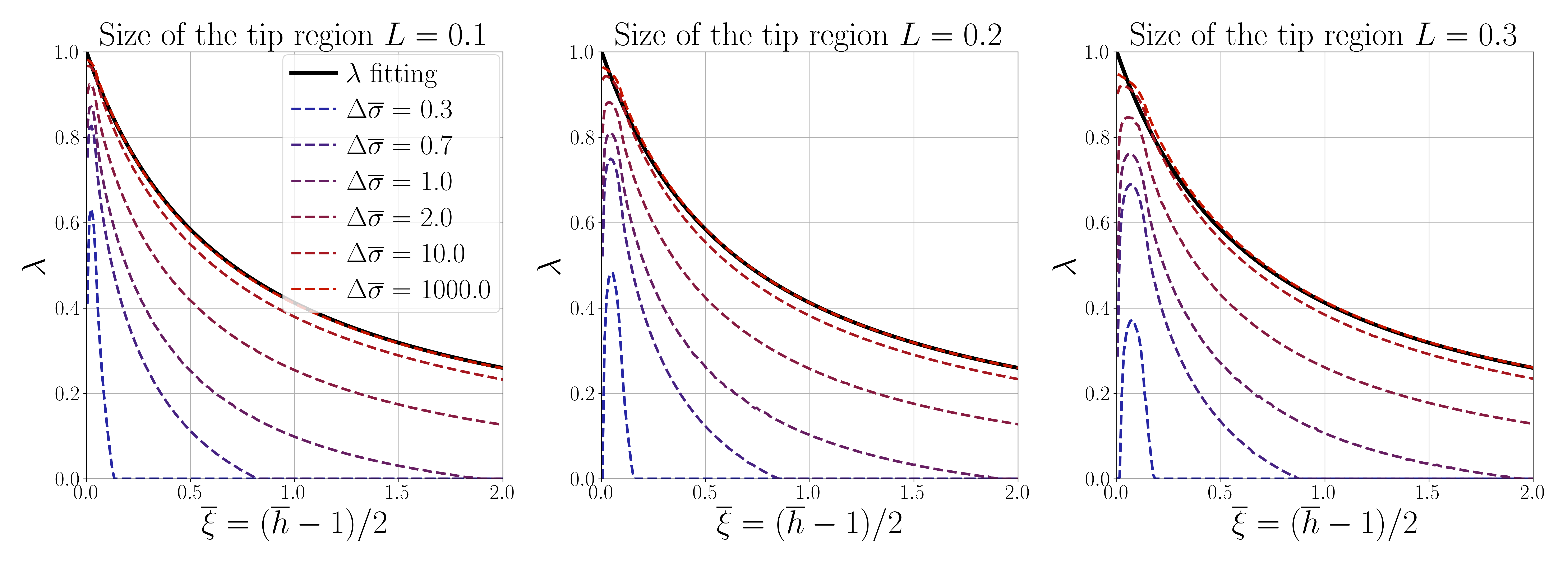}
        \caption{Stress relaxation factor minimizing the average difference between finite fracture solution~\eqref{validity_finite_fracture_width} and the stress-corrected asymptote with the stress relaxation factor~\eqref{validity_stress_relaxation_asymptote} for different magnitudes of the fracture height $\widebar{h}$, the amplitudes of the stress barrier $\Delta\widebar{\sigma}$, and the sizes of the tip region $L$. The black solid line shows the fitting~\eqref{lambda_approximation_fit} of the stress relaxation factor obtained for $\Delta\widebar{\sigma} = 1000$.}
        \label{img:lambda_optimized}
    \end{figure}

    Figure~\ref{img:lambda_optimized} shows the stress relaxation factor computed by~\eqref{lambda_minimization} against the distance $\widebar{\xi}$ from the stress barrier to the fracture tip for different amplitudes of the stress barrier $\Delta\widebar{\sigma}$ and the sizes of the tip region $L$. With the increase in the amplitude of the stress barrier, the stress relaxation factor converges to the solution that does not depend on the size of the tip region $L$. Increasing the size of the tip region $L$ considerably affects the stress relaxation factor only for small amplitudes of the stress barrier $\Delta\widebar{\sigma}$. For this reason, to determine the stress relaxation factor that approximates the solution of the minimization problem~\eqref{lambda_minimization} for different values of the fracture height $\widebar{h}$, the amplitude of the stress barrier $\Delta\widebar{\sigma}$, and the size of the tip region $L$, we fit the curve for $\Delta\widebar{\sigma} = 1000$ that practically does not depend on the size of the tip region to yield
    \begin{equation}\label{lambda_approximation_fit}
        \lambda(\widebar{\xi}) = \frac{0.7}{0.7 + \widebar{\xi}}.
    \end{equation}
    The obtained hyperbolic fitting~\eqref{lambda_approximation_fit} is shown in Figure~\ref{img:lambda_optimized} by the solid black line. Note that for the dimensional cases (particularly $H \neq 1$), the stress relaxation factor depends on the height of the reservoir layer as $\lambda(\xi) = 0.7 / (0.7 + (\xi / H))$. The primary issue with this approach is that the height of the reservoir layer is not well-defined for an arbitrary configuration of the reservoir layers. This problem, despite being very important, is beyond the scope of this study and will be addressed in future studies.

    \begin{figure}
        \centering
        \includegraphics[width=0.99\linewidth]{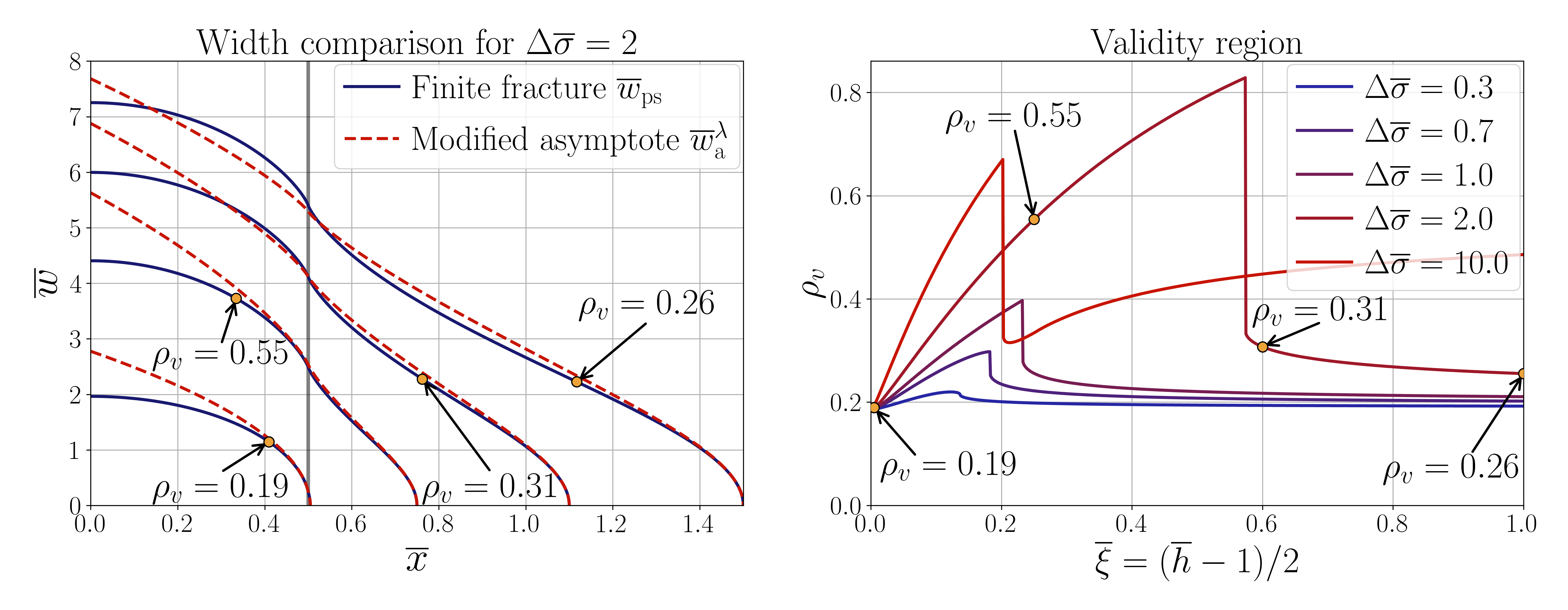}
        \caption{Comparison between the finite fracture solution~\eqref{validity_finite_fracture_width} and the stress-corrected asymptote~\eqref{validity_stress_relaxation_asymptote} with optimal stress relaxation factor~\eqref{lambda_approximation_fit} for several values of fracture height (left). The yellow markers indicate the validity region. The right panel shows the variation of the validity region for the stress-corrected asymptote with the optimal stress relaxation factor versus the distance from the fracture tip to the layer for different amplitudes of the stress barrier $\Delta\widebar{\sigma}$.}
        \label{img:validity_region_corrected}
    \end{figure}

    Figure~\ref{img:validity_region_corrected} depicts the validity region of the stress-corrected asymptote~\eqref{validity_stress_relaxation_asymptote} with optimal stress relaxation factor~\eqref{lambda_approximation_fit}. The left panel in Figure~\ref{img:validity_region_corrected} shows the comparison of the finite fracture width~\eqref{validity_finite_fracture_width} and the modified stress-corrected asymptote~\eqref{validity_stress_relaxation_asymptote} for the fixed amplitude of the stress barrier $\Delta\widebar{\sigma} = 2$. The stress-corrected asymptote with optimal stress relaxation factor matches the finite fracture width much better than the stress-corrected asymptote without stress relaxation factor. The right panel in Figure~\ref{img:validity_region_corrected} shows the validity region of the modified stress-corrected asymptote, which stays above 19 percent for any fracture height and amplitude of the stress barrier. An increase in the validity region, followed by a sharp decline, especially for large values of the amplitude of the stress barrier $\Delta\widebar{\sigma}$, is caused by the non-monotonic behavior of the relative error $|(\widebar{w}_{\mathrm{ps}} - \widebar{w}_{\mathrm{a}}^\lambda) / \widebar{w}_{\mathrm{ps}}|$. In any case, the introduced stress relaxation factor that effectively reduces the amplitude of the stress barrier significantly extends the validity region of the stress-corrected asymptote. And therefore, it is possible to use the modified stress-corrected asymptote along with a relatively coarse mesh to obtain an accurate result and to solve the ``horns'' problem outlined at the beginning of this paper.

\section{Conclusions}
    This paper investigates the problem of a semi-infinite hydraulic fracture propagating in a permeable elastic medium with stress layers. We show that the often-used universal asymptotic solution  may lead to significant errors when the fracture front crosses a stress layer since the effects of the layer are not included. An algorithm that is capable to capture the effects of the stress layer and the multiscale nature of the tip region solution is developed. In particular, we propose three various approaches to incorporate the effect of stress layers. These approaches differ in computational complexity, the complexity of implementation, and the accuracy of the approximation. The first approach involves solving the non-singular integral formulation for the problem. The second, toughness-corrected approach, utilizes the universal asymptotic solution and the concept of an effective toughness, which is calculated using the toughness-dominated asymptote and stress intensity factor correction due to stress layers. Finally, the third approach is based on the ordinary differential equation approximation of the non-singular integral formulation.

    The comparison of the described solutions shows that the integral equation approach has high accuracy, but at the same time, it is computationally inefficient since it requires solving the system of nonlinear equations. The primary advantage of the toughness-corrected approach is the simplicity of implementation and computational efficiency. However, as the influence of viscous dissipation increases, the accuracy of the solution in the presence of stress layers is significantly reduced. The ODE approximation of the integral formulation allows us to capture the influence of stress layers in the case of non-negligible viscous dissipation and leak-off. At the same time, the method is computationally more efficient than the integral equation approach. Thus, the ODE approximation is a better candidate to be implemented into the front tracking algorithm.

    In addition, we evaluate the size of the validity region of the stress-corrected asymptote in the case of a uniformly pressurized plane-strain hydraulic fracture. Results demonstrate that the validity region of the stress-corrected tip asymptote decreases significantly when the fracture front is away from the layer boundary. To increase the validity region of the stress-corrected asymptote, we introduce the stress relaxation factor that effectively reduces the magnitude of the stress barrier. All the modifications presented in this study allow us to increase the accuracy of the layer-crossing computation in advanced hydraulic fracturing simulators, such as ILSA, that utilize the tip asymptotic solutions and to avoid numerical artifacts such as ``horns'' that diminish the overall accuracy of the front tracking.

\section*{Acknowledgements}
    The authors would like to thank Dr. S.V. Golovin and A.N. Baykin for useful discussions at the inception of this project.

\bibliographystyle{elsarticle-num} 
\bibliography{bibliography.bib}
            
\end{document}